\documentclass[superscriptaddress,nobibnotes,amsmath,amssymb,notitlepage,twocolumn,pra,longbibliography]{revtex4-2}

\usepackage{bm,braket}
\usepackage[toc,page]{appendix}
\usepackage{comment}
\usepackage{dcolumn}
\usepackage{multirow}
\usepackage[export]{adjustbox}
\usepackage{amsfonts}

\usepackage{graphicx,color,hyperref}
\usepackage[caption=false]{subfig}
\hypersetup{colorlinks=true, linkcolor=blue, citecolor=blue, urlcolor=blue} 

\graphicspath{{./img/}}

\newcommand{\beq}[1]{\begin{equation}\label{#1}}
\newcommand{\eep}{\;.\end{equation}}
\newcommand{\eec}{\;,\end{equation}}
\newcommand{\eeq}{\end{equation}}

\newcommand*\dd{\mathop{}\!\mathrm{d}} 

\newcommand{\lb}{\left(}
\newcommand{\rb}{\right)}

\renewcommand{\a}{\alpha}
\renewcommand{\b}{\beta}
\newcommand{\g}{\gamma}
\renewcommand{\d}{\delta}
\newcommand{\ep}{\epsilon}

\newcommand{\la}{\lambda}
\renewcommand{\th}{\theta}
\newcommand{\s}{\sigma}
\newcommand{\p}{\phi}
\newcommand{\om}{\omega}

\newcommand{\D}{\Delta}
\newcommand{\G}{\Gamma}

\newcommand{\Om}{\Omega}
\newcommand{\La}{\Lambda}
\DeclareMathAlphabet{\mathcal}{OMS}{cmsy}{m}{n} 

\newcommand{\Df}{\mathcal{D}}    

\newcommand{\bigO}{O} 

\renewcommand{\vec}[1]{{\bf #1}}

\newcommand{\kv}{\vec{k}}

\newcommand{\vp}{\vec{p}}

\newcommand{\A}{\vec{A}}

\usepackage{amsmath}
\usepackage{amssymb}
\usepackage{xcolor}
\usepackage{bbm}
\usepackage{physics}
\usepackage{float}
\usepackage{graphicx}
\usepackage{dcolumn} 
\usepackage{bm} 
\usepackage{siunitx}
\usepackage{enumitem}  

\makeatletter
\renewcommand*{\fnum@figure}{{\normalfont\bfseries \figurename~\thefigure}}
\makeatother

\allowdisplaybreaks

\definecolor{orange}{rgb}{1,0.5,0}

\newcommand{\figref}[1]{Fig.~\ref{#1}}

\renewcommand{\a}{\alpha}
\renewcommand{\b}{\beta}
\renewcommand{\d}{\delta}

\renewcommand{\th}{\theta}

\DeclareMathAlphabet{\mathcal}{OMS}{cmsy}{m}{n} 

\usepackage{pifont}
\newcommand{\cmark}{\ding{51}}%
\newcommand{\xmark}{\ding{55}}%

\newcommand{\ra}{\rightarrow}
\newcommand{\Ra}{\Rightarrow}
\newcommand{\Rra}{\Rrightarrow}
\newcommand{\HBdG}{H_\mathrm{BdG}}
\newcommand{\ept}{\tilde{\ep}}

\renewcommand{\Re}{\mathfrak{Re}~}

\makeatletter
\newcommand{\specificthanks}[1]{\@fnsymbol{#1}}
\makeatother

\begin{document}

\title{Optical signatures of Euler superconductors}

\author{Chun Wang Chau}
\email{cwc61@cam.ac.uk}
\affiliation{TCM Group, Cavendish Laboratory, Department of Physics, J J Thomson Avenue, Cambridge CB3 0HE, United Kingdom}

\author{Wojciech J. Jankowski}
\affiliation{TCM Group, Cavendish Laboratory, Department of Physics, J J Thomson Avenue, Cambridge CB3 0HE, United Kingdom}

\author{Robert-Jan Slager}
\email{rjs269@cam.ac.uk}
\affiliation{Department of Physics and Astronomy, University of Manchester, Oxford Road, Manchester M13 9PL, United Kingdom}
\affiliation{TCM Group, Cavendish Laboratory, Department of Physics, J J Thomson Avenue, Cambridge CB3 0HE, United Kingdom}
\date{\today}

\begin{abstract}
    We study optical manifestations of multigap band topology in multiband superconductors with a~nontrivial topological Euler class. We introduce a set of lattice models for non-Abelian superconductors with the Euler invariant signified by a nontrivial quantum geometry. We then demonstrate that the topological Bogoliubov excitations realized in these models provide for a characteristic first-order optical response distinct from those of the other known topological superconductors. We find that the spectral distribution of the optical conductivity universally admits a topological jump originating from the Euler class in the presence of $d$-wave superconducting pairings, and naturally differs from the features induced by the quantum geometry in the noninteracting bands without pairing terms. Further to uncovering observable signatures in first-order optical conductivities, we showcase that the higher-order optical responses of the non-Abelian Euler superconductor can result in enhanced nonlinear currents that fingerprint the exotic topological invariant. Finally, by employing a diagrammatic approach, we generalize our findings beyond the specific models of Euler superconductors.
\end{abstract}

\maketitle

\section{Introduction}
The study of topology in condensed matter systems, in particular, topological insulators, semimetals, and superconductors, has been an active field for the last few decades \cite{Rmp1,Rmp2,Rmp3}. While experimentally challenging~\cite{Lutchyn2010, Mourik_2012, Sarma_2015, Zhu_2020, Yazdani_2023}, from a~theoretical perspective, 
topological superconductors, in particular, host the intriguing possibility of being a platform for Majorana excitations~\cite{Kitaev2001, Fu2008, Beenakker2013}. Their interplay with topological insulators in engineered device setups, is of central interest, as these offer numerous promises to realize exotic proximity effects~\cite{Fu2008}. Last but not least, emergent Majorana zero modes are predicted to be usable for fault-tolerant topological quantum computation~\cite{Kitaev2006, Rmp4}. 

Although single-gap topological insulators and topological semimetals are fairly extensively classified~\cite{Ludwig, Kitaev2009, Slager_2013, Shiozaki14, Kruthoff_2017, Slager2019, Po_2017, Bradlyn_2017}, recently, additional multigap topological phases have been discovered~\cite{bouhon2020geometric}. In these systems, groups of bands, or band subspaces, carry previously uncharted multigap invariants~\cite{bouhon2020geometric, davoyan2024, Bouhon2024}. A prominent example of a multigap invariant is the Euler class $\chi$~\cite{bouhon2018wilson, BJY_nielsen, bouhon2020geometric}, which characterizes systems described by real-valued Hamiltonians due to the presence of $\mathcal{C}_2\mathcal{T}$, i.e., twofold rotation combined with time reversal, or $\mathcal{PT}$, i.e. inversion combined with time reversal symmetry. In such systems, the nontrivial Euler class can be associated with non-Abelian band singularities, which we refer to as Euler nodes. An Euler node characterized with an Euler class charge $\chi$ can be split into $2|\chi|$ nodes of the same chirality, and a pair of bands with an Euler class $\chi$ admits $2|\chi|$ topologically protected nodes, consistently with the Poincar\'e-Hopf theorem~\cite{bouhon2020geometric}. Distinctively from Weyl or multi-Weyl nodes, Euler nodes are not monopoles of Berry curvature and instead act purely as sources of quantum metric over the momentum space~\cite{Bouhon2023geometric, jankowski2023optical}. As relevant to the rest of this work, splitting a quadratic Euler node results in a formation of two linear Dirac or Weyl nodes of the same chirality, depending on whether the $\mathcal{P}\mathcal{T}$ symmetry is broken or preserved. If the symmetry is preserved, the individual nodes formed on splitting can be moreover characterized with a half-integer patch Euler class~\cite{Siyu2025}. Mathematically, the band nodes are characterized by the same homotopy relations as the $\pi$-disclination defects in biaxial nematics~\cite{Wu1273,volovik2018investigation,Genqcs2016,Beekman20171,bouhon2019nonabelian, Jiang2021, Kamienrmp}. Such band degeneracies carry non-Abelian quaternion frame charges, which can be manipulated by braiding nodes between multiple gaps in the momentum space~\cite{BJY_nielsen, bouhon2019nonabelian, bouhon2022multigap}. As a result, momentum-space nodal braiding allows us to have nodes with the same charges within a two-band subspace. This ensures the Euler class attains a finite value, which pinpoints the pairs of nodes that can be merged but cannot be annihilated. Their charges can otherwise be changed upon braiding with a band node in an adjacent band gap, which ensures a topological protection as long as the gap is maintained~\cite{bouhon2019nonabelian, BJY_nielsen, Jiang2021, Jankowski2024disorder}. 

\begin{figure*}[t]
    \includegraphics[width=0.96\linewidth]{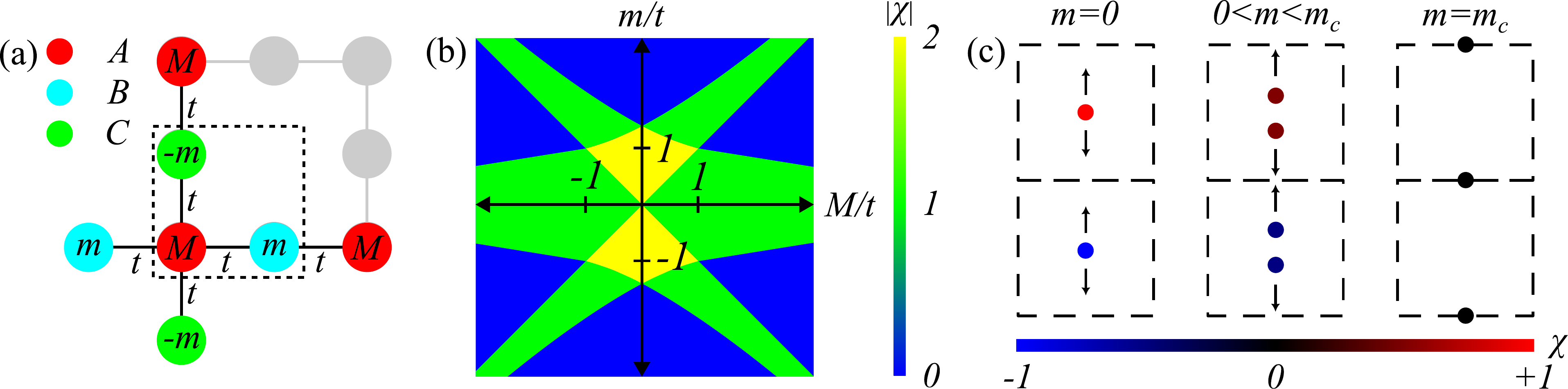}
    \caption{Realizing nontrivial Euler class on the Lieb lattice. {\bf (a)} Lattice structure of the tight-binding model. {\bf (b)} Phase diagram realized in the model as the on-site energy parameters $M$ and $m$ are changed. The phase is classified by the total patch Euler number of all nodes $\sum_n\vert\chi_n\vert$. {\bf (c)} Example of splitting of the Euler node as we increase $m$. We split a single node of patch Euler class $\chi = 1$ located at the $K$ point [$(k_x, k_y) = (\pi, \pi)$] into two nodes of patch Euler class $\chi = 1/2$, which move in opposite directions toward the $Y$ point [$(k_x, k_y) = (0, \pi)$] or toward the $X$ point [$(k_x, k_y) = (\pi, 0)$] if we decrease $m$ instead. For $m=m_c$, the nodes annihilate at the BZ edge on merging with their partners from the adjacent BZ, which results in a vanishing total Euler class $\chi = 0$. We note that one Euler node realizes a positive and the other one a negative charge on merging, as the Euler bands in the studied model are unorientable, and thus the charges in the neighboring BZs are opposite~\cite{Jiang2021,jiang_meron}.}
    \label{fig1}
\end{figure*}

Rather than remaining a purely theoretical pursuit, multigap invariants are increasingly retrieved in a wide range of systems. To date, they have been observed in trapped-ion experiments~\cite{unal2020, zhao2022observation} and metamaterials~\cite{Jiang2021, jiang_meron}, with additional directions including electron~\cite{bouhon2019nonabelian, lee2024,PhysRevB.103.245127} and phonon band topologies~\cite{Peng2021, Peng2022Multi} of multiple materials simulated with first-principles calculations. Moreover, in the context of superconductors, we point out that recently the Euler class has also been predicted to induce an obstruction in the formation of Cooper pairs, which was suggested in the context of twisted graphene bilayers~\cite{yu2022,yu2023}. A physical smoking-gun signature of the Euler class to experimentally validate these predictions in real materials therefore remains a central topic of interest. In this context, the previous studies considered optical manifestations, such as optical conductivities or higher-order bulk photovoltaic effects~\cite{jankowski2023optical,jankowskiPRL2024, jankowskiPRB2024Hopf}. Similarly, transport signatures, including linear and nonlinear anomalous Hall responses~\cite{haldane1988, Gao2014, Gao_2023_2}, were predicted to probe the topology of multigap topological insulators and semimetals~\cite{jain2024}. In the case of topological superconductors included in the tenfold classification~\cite{Kitaev2009}, the optical signatures were previously identified~\cite{he2021}, while in the multigap case only the scattering signatures, such as Andreev reflections~\cite{Morris_2024}, have been discussed. In contrast, the optical responses of such exotic phases remain an open problem, which we address in this work. In this regard, the notions of quantum geometry~\cite{provost1980riemannian, Peotta_2015, Torma2023, Bouhon2023geometric} and their connections to the optical responses~\cite{Ahn2020, Ahn2021, Bouhon2023geometric} are of particular interest for the superconducting states, given that quantum geometry naturally indicates the presence of nontrivial multigap invariants~\cite{Bouhon2023geometric, jankowski2023optical}, such as the Euler class. The importance and role of quantum geometry for the optical responses of superconducting states has most recently gained increasing recognition~\cite{AhnSC2021, Chen2021, watanabe2022, Tanaka2023nonlinear, tanaka2024nonlinear, watanabe2024, Kotetes2024}.

\begin{figure*}[t]
    \raggedleft
    \includegraphics[width=1.0\linewidth, right]{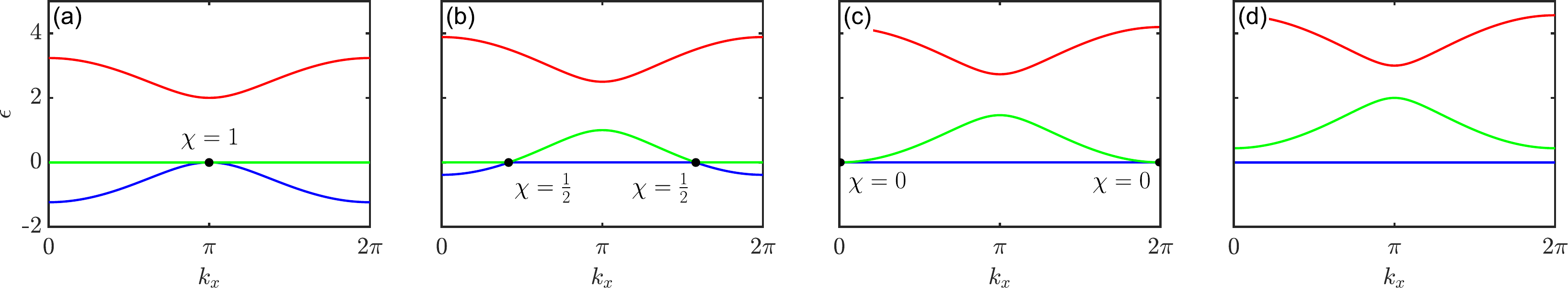}
    \caption{Evolution of the band structure and Euler node of the Lieb lattice along $k_y=\pi$. {\bf (a)} Band structure of the Euler Hamiltonian [Eq.~\eqref{HLieb}] with ${t=1},\;{M=2}$ and $m=0$ including all bands. The pair of Euler bands are separated from a higher band by mass gap $M$ and an integer Euler node of $\chi=1$ is located at the $K$ point. {\bf (b)} Bands at $m=0.5$, at which the integer Euler node is split into two half integer Euler nodes that can be associated with a patch Euler class $\chi=1/2$ individually, as we increase $m$. {\bf (c)} Bands at the critical $m=m_c\sim0.73$, at which pair annihilation of half-integer Euler nodes occurs, resulting in trivial node at $(k_x,k_y)=(0,\pi)$, as demonstrated in \figref{fig1}(c). {\bf (d)}~Increasing $m$ further to $m=1$ results in a trivial multigap phase, where all bands are gapped, which results in vanishing patch Euler class $\chi$.}
    \label{figband}
\end{figure*}

In this work, we investigate the optical response in superconductors realizing a nontrivial Euler class invariant using concrete lattice models and a model-independent diagrammatic approach. We obtain signatures in the linear optical conductivity and higher-order photoconductivities unique to superconducting systems, which we further elucidate within the framework of quantum geometry. The unique features of the Euler superconductors retrieved in this work involve a signature linear optical conductivity jump at the frequency corresponding to the quasiparticle photoexcitation across the superconducting gap to the Euler node carrying the integer patch Euler class between the dispersive and flat bands. Moreover, the quantum metric bounds due to the Euler curvature present in the Euler superconductors provide for an enhancement of nonlinear optical responses captured by the third-order photoconductivity tensors. The uniqueness of these features arises, respectively, from the singular and topologically lower-bounded quantum geometries induced by the Euler invariant, which we detail in Appendix~\ref{app::A}. To expose these optical features, we more specifically start from a tight-binding model that hosts Euler bands on a Lieb lattice, then introduce intraband interactions, i.e., pairing terms, and construct a Bogoliubov-de-Gennes~(BdG) Hamiltonian to describe a topological Euler superconductor at the mean-field level. 
We find that in the superconducting state, tuning the phase difference between order parameters of the Euler bands can enhance the optical transition between Euler nodes of different sectors, i.e., particle/hole ($\pm$) sectors, in the BdG Hamiltonian spectrum. Furthermore, we retrieve a signature jump originating from the
Euler class in the real part of the linear optical conductivity in the presence of a $d$-wave superconducting pairing, which is proportional to the Euler invariant and is induced by the singular quantum geometry realized in the Euler superconductor.
We furthermore investigate noncentrosymmetric Euler superconductors, where both $\mathcal{P}$ and $\mathcal{T}$ symmetry are broken, which allows for nonvanishing responses at second order in optical fields. We then observe that the response is dominated by the injection currents, which are governed by the multiband quantum metric. 
However, we note that the intersector quantum metric, i.e., quantum metric combining states of both particle and hole sectors, exactly at the Euler node, is vanishing. Thus, no direct topological, e.g., quantized, signature of the Euler class can be observed in the second-order response, unlike in the first-order response. We further find that beyond the first-order optical response, there are topologically enhanced diagrammatic contributions at third order in optical electric fields.
In that context, we construct additional selection rules that arise from the difference in the coupling of electrons and holes with electromagnetic fields. We show that these could forbid certain third-order responses manifested in the normal state but otherwise introduce signatures that are unique to the superconducting system.
\\

The paper is organized as follows. In Sec.~\ref{sec::II}, we provide details on realizations of Euler multiband superconductors adapted to lattice tight-binding models. In Sec.~\ref{sec::III}, we detail the superconducting pairings and their symmetries in the considered Euler superconductor models. In Secs.~\ref{sec::IV} and~\ref{sec::V}, we correspondingly employ linear and nonlinear response theories to numerically benchmark the optical responses at linear and nonlinear orders. Sec.~\ref{sec::VI} then provides the diagrammatic analysis of the linear and nonlinear responses of non-Abelian Euler superconductors beyond the introduced specific models. In this context, we retrieve Euler class-induced topological contributions within one-loop diagrams for first- and third-order responses. We further discuss these findings in Sec.~\ref{sec::VII}, before concluding and providing an outlook on the experimental feasibility of measuring the Euler superconductor invariant using optical probes, as applied to physical materials, in Sec.~\ref{sec::VIII}. 
\begin{figure*}[t]
    \centering
    \includegraphics[width=1\linewidth]{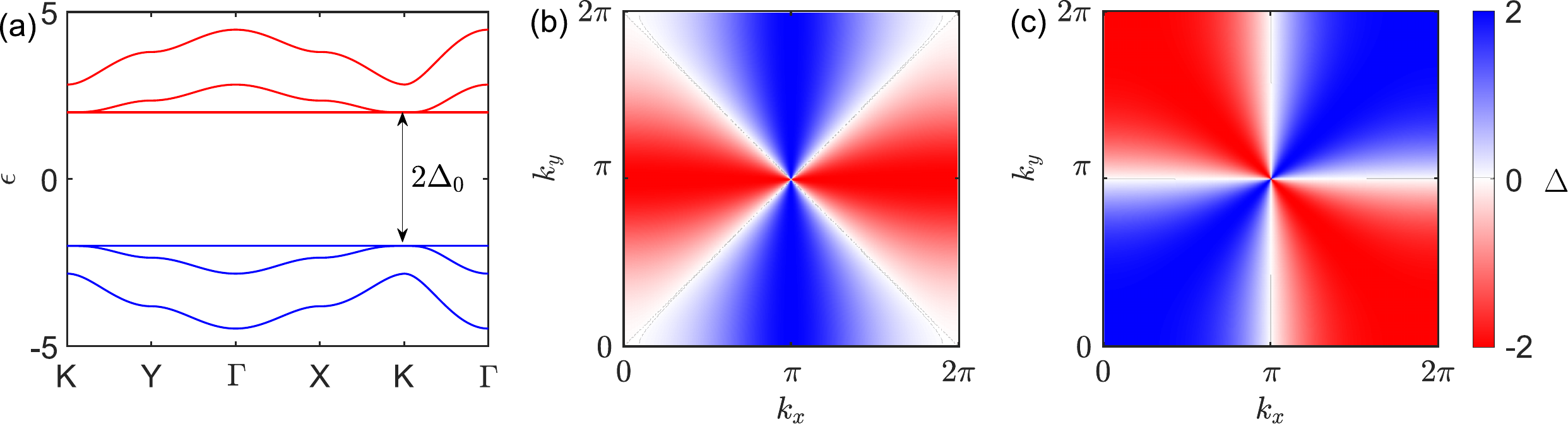}
    \caption{Euler superconductor on the Lieb lattice. {\bf (a)} Band structure for the BdG Euler Hamiltonian [Eq.~\eqref{HBdG}] with ${t=1},\;{M=2},\;m=0$, and $\D_0=2$ for all bands. The electron sector (\textit{red}) and hole sector (\textit{blue}) are separated by a gap $2\D_0$. BdG band structures for particle sector generated from different phases in the normal state. Visualization of pairing symmetry in the momentum space when $\p=\pi$, of $\D_{BB}(\kv)$, the pairing between $B$ orbitals {\bf (b)}, and $\D_{BC}(\kv)$ {\bf (c)}, the pairing between $B$ and $C$ orbitals. Both pairings exhibit $d$-wave symmetry, respectively, represented by $d_{x^2-y^2}$ and $d_{xy}$. Thus, $\phi=\pi$ corresponds to a realization of a $d$-wave superconductor within the Lieb lattice.}
    \label{fig2}
\end{figure*}
\section{Lieb Lattice Model}\label{sec::II} 
While most of our results are universal to general systems with a nodal Euler invariant, as we further elucidate within a diagrammatic approach,  we start by considering a model on a Lieb lattice. The model is illustrated in \figref{fig1}(a) and the corresponding Hamiltonian reads 
\beq{HLieb}
        h_0(\kv) =
        \begin{pmatrix}
            M &   2t\cos\frac{k_x}{2} &   2t\cos\frac{k_y}{2} \\
            2t\cos\frac{k_x}{2} &   m   &   0   \\
            2t\cos\frac{k_y}{2} &   0   &   -m   \\
        \end{pmatrix}.
\eeq
 The Hamiltonian in Eq.~\eqref{HLieb} satisfies the reality condition, as ensured by the $\mathcal{PT}$ symmetry~\cite{bouhon2019nonabelian}. For convenience, we consider parameter pairs $(c_{1},c_{2})=(r\cos\th,r\sin\th)=\lb2t\cos\frac{k_{x}}{2},2t\cos\frac{k_{y}}{2}\rb$, both with values ranging from $-2t$ to $2t$. On setting $m=0$, the Hamiltonian $h_{0}(\kv)$ yields Bloch eigenvectors $u_{a}(\kv)$:
    \begin{align}
        u_f&=
        \begin{pmatrix}
            0\\
            -\sin\th\\
            \cos\th
        \end{pmatrix}\;,\\
        u_\pm &=
        \frac{1}{\sqrt{r^2+\ep_\pm^{2}}}
        \begin{pmatrix}
            \ep_\pm\\
            r\cos\th\\
            r\sin\th
        \end{pmatrix}\;,
    \end{align}
where $\ep_\pm=\frac{M}{2}\pm\sqrt{r^2+\frac{M^{2}}{4}}$ represents the energy spectrum of the dispersive band.
The model realizes a flat band located at energy $\ep_f = 0$, and has a quadratic band node between the flat band and dispersive band with energy $\ep_-$ ($\ep_+$) for $M>0$ ($M<0$). The node realizes nontrivial band topology and quantum geometry associated with the Euler class invariant (see Appendix~\ref{app::A}). The Euler class $\chi_{ab}$ over a two-dimensional Brillouin zone (BZ) patch $\Df$ is defined as a topological obstruction to Stokes' theorem~\cite{BJY_nielsen, bouhon2019nonabelian}:
\beq{Eu_Patch}
    \chi_{ab}(\Df) = 
    \frac{1}{2\pi}\left[
        \int_\Df [d\kv]~\mathrm{Eu}_{ab}(\kv) 
        -
        \oint_{\partial\Df} d\kv \cdot \mathbf{\tilde{\xi}}_{ab}
    \right]
\eec
which captures the singular non-Abelian multiband connection between bands $a$ and $b$, $\tilde{\xi}^\mu_{ab} = \langle a \vert \partial^\mu b \rangle$, which determines the Euler curvature, $\mathrm{Eu}_{ab} = \nabla_\kv \times \tilde{\xi}_{ab}$; see Appendix~\ref{app::A} for more details. $\partial^\mu$ denotes a partial derivative in crystal momentum component $k_\mu$, with $\mu = x,~y$. The Euler class $\chi$ defined on a patch between selected two bands $a,b$ corresponds to an element of the matrix Euler class $\chi_{ab}$.
The invariant is nonvanishing and quantized $[\chi \in \mathbb{Z}~(+\frac{1}{2})]$~\cite{BJY_nielsen, bouhon2019nonabelian, Siyu2025}, if and only if the Hamiltonian is invariant under the action of $C_2\mathcal{T}$ or $\mathcal{PT}$ symmetry. One can verify that the patch Euler class $\chi$ for the band node between the flat band and dispersive band amounts to $\chi = 1$. 
\begin{figure*}[t]
    \centering
    \includegraphics[width=1\linewidth]{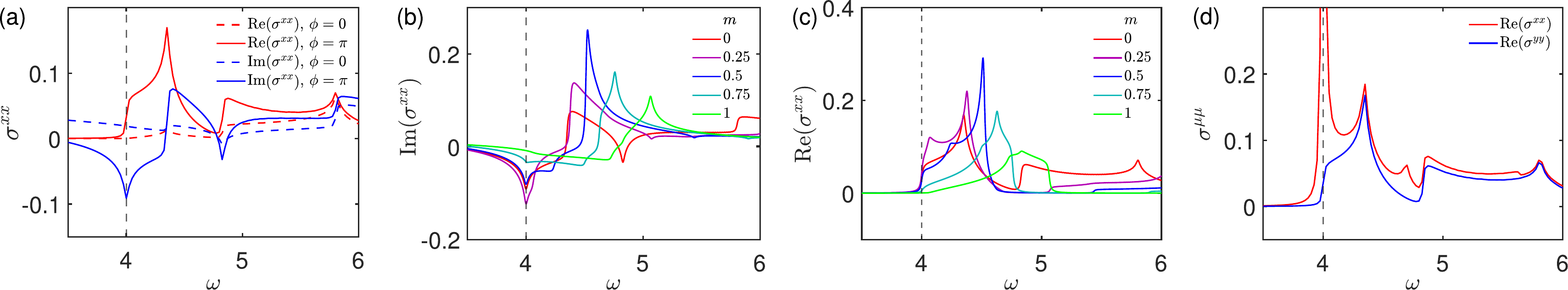}
    \caption{First-order photoconductivity, with chemical potential exactly at the Euler node, and $M=2,\;t=1,\;\D_0=2$. We use a vertical dashed line to denote the superconducting gap energy $2\D_0$ {\bf (a)} Real part of conductivity (\textit{red}) and imaginary part of $\s^{xx}$ (\textit{blue}) for $m=0$. Notably, $\s^{xx} = \s^{yy}$ and $\s^{xy}=0$ due to vanishing integral of multiband metric $g^{xy}_{ab}$ over the Brillouin zone. We further compare the cases with zero phase difference $\phi=0$ in the order parameter (\textit{dashed line}) and $\phi = \pi$ phase difference (\textit{solid line}). For the real part, there is a discrete jump of $e^2/(16\hbar)$ as a manifestation of the Euler class, as given by Eq.~\eqref{rPC}. When the phase difference is vanishing, the conductivity jump is suppressed. {\bf (b)} Imaginary part and {\bf (c)} real part of $\s^{xx}$ as we vary $m$. The Euler nodes are annihilated when $m =0.73$, thus a discrete jump cannot be observed for $m=0.75,\;1$. {\bf (d)} We introduce a $\mathcal{P}$-breaking term $t_2\sin{k_x}$ along the $x$ direction to all diagonal terms of the Hamiltonian and set $t_2 = 0.5$. We note that the signature remains unchanged for $\s^{yy}$, as the $y$ direction respects the inversion symmetry, whereas the discrete jump evolves into a singular peak for $\s^{xx}$.}
    \label{fig3}
\end{figure*}
A complete phase diagram of the model, as a function of $M,\;m$, is given in \figref{fig1}(b). 
We also illustrate the splitting of the Euler node of $\chi=1$, into two nodes which act as half vortices in the multiband connection (${\chi=1/2}$) as we increase $m$ from $m=0$ in \figref{fig1}(c). For ${M>0}$, we note that the split nodes evolve and annihilate at the BZ edge as we increase $m$ to a critical value $m_c$, which results in a trivial phase via gap opening.
The band structure details for different phases, as we change $m$, are illustrated in \figref{figband}. Respectively, we present the splitting of the integer Euler node [\figref{figband}(a)] into two half-integer Euler nodes [\figref{figband}(b)] as we increase $m$, which results in topologically trivial system [\figref{figband}(d)] when $m$ is increased further, as the pair annihilation of the half-integer Euler nodes occurs when $m=m_c$ [\figref{figband}(c)].

We now provide details on the superconducting phase supported by the introduced model in the presence of pairing interactions, as our main aim is to address the bulk physical signatures of the superconductor with nontrivial Euler class. At the mean-field level, we can describe the superconducting phase by introducing the BdG Hamiltonian:
\beq{HBdG}
    H(\kv)
    =
    \left.
    \begin{pmatrix}
        h_0(\kv-\A)   &   \D(\kv) \\
        \D^\dagger(\kv) &   -h_0(-\kv-\A)^T   \\
    \end{pmatrix}
    \right\vert_{\A=0}
\eec
where we have assumed the order parameter does not couple to, or depend on, the vector gauge field $\vec{A}$. On restricting the superconductor pairing terms to intraband interactions (see Sec.~\ref{sec::III}), the order parameter can be decomposed as
\beq{Dk}
    \D(\kv) = \sum_a \D_a \ket{u_a(\kv)}\bra{u_a(\kv)}
\eec
where $u_a(\kv)$ are the Bloch vectors of the normal state, and $\D_a$ denote the order parameters in bands $a = \pm, f$. The eigenstates $\tilde{u}_{a,s}$ of the BdG Hamiltonian thus can be related to the eigenstates of the normal state:
\begin{align}
    \tilde{u}_{a,s} &= \frac{1}{\sqrt{1+|\a_{a,s}|^2}}
    \begin{pmatrix}
        \a_{a,s}    \\  1    
    \end{pmatrix}
    \otimes
    u_a\;,
    \\
    \a_{a,\pm} &= \frac{\ep_a\pm\sqrt{|\D_a|^2+\ep_a^2}}{\D_a^*} \;, \\
    \ept_{a,\pm} &= \pm\sqrt{|\D_a|^2+\ep_a^2} \;,
\end{align}
where eigenstates of the BdG Hamiltonian $\vert \tilde{u}_{a,s} \rangle$ are indexed by $a$, i.e., the normal band index, and by $s=\pm 1$, the sector index, with $+1\;(-1)$ corresponding to the electron (hole) sector. The spectrum of the quasiparticle bands in the Euler superconductor is demonstrated in \figref{fig2}(a). In the following, we discuss the superconducting pairing and its symmetries in the considered model.
\section{Pairing symmetry in the Euler superconductor}\label{sec::III}

Further to introducing the Lieb lattice Euler superconductor model, we elaborate on the relevant pairing symmetries consistent with the considered lattice Hamiltonian. In particular, we discuss the relationship between the order parameter phase difference $\phi$ in the band basis and the nontrivial pairing in the orbital basis.
We begin with the trivial coupling case of $\p=0$, which can be connected to the case $\D(\kv)=\D_0 \mathbbm{1}_3$, assuming that $\D_+=\D_-=\D_f \equiv \D_0$ corresponds to a uniform $s$-wave pairing in momentum space. On the contrary, without changing the band structure, i.e., keeping $|\D_a| = \D_0$ for all bands, by setting the phase difference to $\p=\pi$ between $\D_-$ and $\D_f$ ($\D_-=e^{\text{i}\phi} \D_f = -\D_f$), with $\D_+=\D_-=-\D_0$, we have a multiband order \text{parameter}:
\begin{align}
    \D(\p=\pi)
    &=
    -\D_0
    \begin{pmatrix}
        1   &   0   &   0   \\
        0   &   \cos2\th    &   \sin2\th    \\
        0   &   \sin2\th    &   -\cos2\th   \\
    \end{pmatrix}
    \nonumber
    \\
    &=
    -\D_0\oplus\frac{1}{r^2}(-\D_{x^2-y^2}\tau_z-2\D_{xy}\tau_x)\;,
\end{align}
where we have defined $d$-wave order parameters $\D_{x^2-y^2}=\D_0r^2\cos2\th$ and $\D_{xy}=\D_0r^2\sin\th\cos{\th}$.
In the orbital basis, the order parameter can be written as a tensor sum of the $s$-wave in the $A$ orbital, together with the $d$-wave in $B$ and $C$ orbitals. As such, when we change the phase difference $\phi$ within the order parameter, we are effectively considering nontrivial pairing in the orbital basis, although there is no change in the quasiparticle band dispersion spectrum. The $d$-wave pairing symmetries between different orbitals are visualized in \figref{fig2}(b) and \figref{fig2}(c). In the following, we will be particularly interested in comparing the case between $\p=0$ and $\p=\pi$, namely, uniform conventional $s$-wave pairing, and nontrivial $d$-wave pairing, which is reminiscent of unconventional $d$-wave superconductivity observed, for example, in Lieb lattice cuprates~\cite{RMPCuprate}. More directly, however, our model is relevant to synthetic matter realizations of Euler topology~\cite{zhao2022observation, unal2020,Jiang2021, jiang_meron}. Having introduced the lattice model and relevant pairing terms for an Euler superconductor, we now study the optical responses induced by the topological multiband invariant in the superconducting state.

\section{Linear optical conductivity}\label{sec::IV} 
We now address the linear optical conductivity in the introduced Euler superconductor.
The linear optical conductivity captures ac current densities $j^\mu(\om)$ in response to the optical electric field $\mathbf{E}(\om) = \frac{1}{2}(\mathbf{E}_0 e^{i \om t} + \mathbf{E}^*_0 e^{-i \om t})$, with a frequency $\om$ and field strength $|\mathbf{E}_0|$, as
\beq{}
 j^\mu(\om) = \sum_{\nu = x, y} \sigma^{\mu \nu}(\om) E_\nu (\om)
\eep
The optical conductivity derived within the linear response theory applied to the superconducting state \cite{watanabe2022} reads
\beq{conlin}
    \s^{\mu\nu}(\om) = \frac{i}{2(\om+i\eta)} \sum_{a,b}\lb
    \frac{J^\mu_{ab}J^\nu_{ba}f_{ab}}{\om+i\eta-E_{ba}}+J^{\mu\nu}_{ab}f_a\d_{ab}\rb
\eec
where $f_{ab} = f_a - f_b$ are the differences in the filling factors, $E_{ba} = \ept_{b,+} - \ept_{a,-}$, and we have set $e=\hbar=1$. Here we also set $\eta \rightarrow 0^+$, and defined the generalized velocity operator for a superconductor (see Appendix~\ref{app::C}) with a vector potential denoted as~$\vec{A}$:
\beq{vo}
    J^{\mu_1\ldots \mu_n}(\kv)
    =\left.(-1)^n\frac{\partial H(\kv,\A)}{\partial A_{\mu_1}\ldots\partial A_{\mu_n}} \right\vert_{\A\ra 0}.
\eeq
We note that the generalized velocity operators can be naturally computed as numerical derivatives. As such, one can easily determine the linear optical conductivity using Eq.~\eqref{conlin} for a specific model; see Appendix~\ref{app::D}. In our case, using the above Lieb lattice model, we can investigate possible optical signatures that encapsulate the Euler class. We, however, comment on a minor but important complication when connecting the generalized velocity operator in quasiparticle band basis back to the velocity operator in the normal state. Namely, because of the difference in coupling of the electron and hole sectors with the vector potential, the optical response associated with the topology of the Euler class originally present in the normal state~\cite{jankowski2023optical} is not directly inherited in the superconducting state.
\begin{figure*}[t]
    \centering
    \includegraphics[width=1\linewidth]{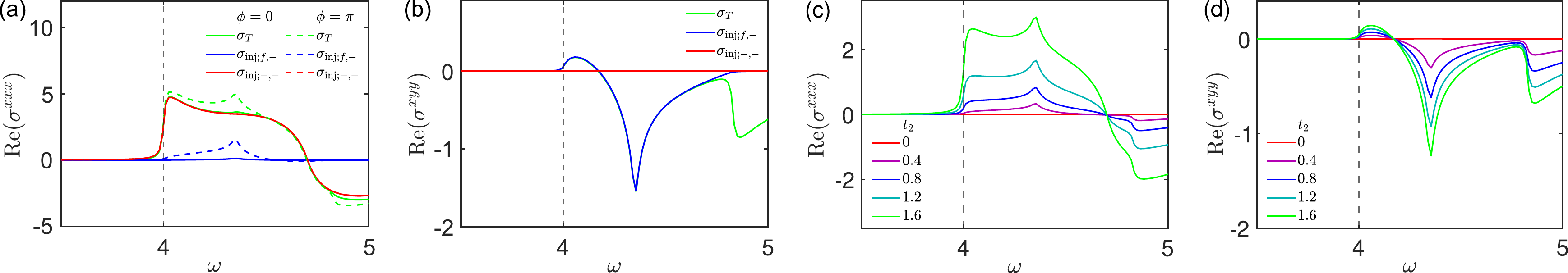}
    \caption{Second-order photoconductivities in a noncentrosymmetric Euler superconductor. {\bf (a)} Real part of $\s^{xxx}$, when $M=2,\;t=1,\;{t_2=2},\;\D_0=2,\;\eta=10^{-2}$, and the injection conductivity contributions from different bands for phase differences $\phi = 0$ and $\phi = \pi$. For both phase differences, comparing with the total conductivity $\s_T$, the main contribution arises from injection conductivity $\s_{\mathrm{inj};-,-}$ that corresponds to the transition between two bands that originate from the same normal state band. While the injection conductivity is proportional to the quantum metric, $\s_{\mathrm{inj};-,-}$, the information about the multiband geometry of the normal state cannot be directly inferred, as discussed in Appendix~\ref{PB}. {\bf (b)} Real part of $\s^{xyy}$. We observe that $\s_{\mathrm{inj};-,-}$ vanishes, as the $y$ direction respects parity. Real part of {\bf (c)} $\s^{xxx}$ and {\bf (d)} $\s^{xyy}$, for different values of $t_2$. We note that the conductivities increase as the parity breaking parameter $t_2$ is increased, with no major change to the spectral profile of the photoconductivities.}
    \label{fig4}
\end{figure*}

In Fig.~\ref{fig3}, we show a numerical computation of the linear conductivity, where we fixed $M=2,\;t=1,\;\D_0=2$. 
To begin, we note a difference in response, as we change the phase difference $\phi$ between the order parameter of the Euler bands (see Sec.~\ref{sec::III}), as illustrated in Fig.~\ref{fig3}(a). We note that for $\p=0$ the response is suppressed. Meanwhile, for $\p=\pi$, at $\om=2\D_0$, which corresponds to the superconducting gap, we note a discrete jump for the real part, and a cusplike profile for the imaginary part. We further show how these optical conductivity features are modified, by varying $m$, which increased beyond the critical value $m_c \sim 0.73$ in our model parametrization, removes the Euler node. In Fig.~\ref{fig3}(b), we observe that the cusp-like profile in the imaginary part of the conductivity flattens as we gap out the Euler node. In contrast, for the real part illustrated in Fig.~\ref{fig3}(c), we note that a discrete conductivity jump persists as we change $m$, as long as the patch Euler class is nonvanishing ($\chi \neq 0$). When the Euler node is gapped out, the discrete conductivity jump is no longer observable, showing a clear sign of transition to the trivial Euler class, contrary to the imaginary part. For completeness, we also consider the case where inversion symmetry $\mathcal{P}$ is broken along the $x$ direction by introducing $\mathcal{P}$ breaking term $t_2\sin{k_x}$, to all diagonal terms of the Hamiltonian while keeping the $\mathcal{PT}$ symmetry intact. We note that upon breaking the $\mathcal{P}$ symmetry along the $y$ direction, the conductivity profile remains unchanged. However, for the $x$ direction, the discrete jump, previously observed at $\om=2\D_0$, evolves into a divergence, as the parity-odd velocity operator is now nonvanishing, 
as illustrated in Fig.~\ref{fig3}(d). To address these numerically retrieved features and their interplay with the Euler class further, we perform a diagrammatic analysis within continuum models in Sec.~\ref{sec::VI}. In the next section, we focus on the numerical results concerning higher-order optical responses.

\section{Higher-order responses}\label{sec::V} 

To address the higher-order responses of an Euler superconductor, we first study the second-order photoconductivities. The second-order photoconductivity $\sigma^{\mu \nu \rho}(\om; \om_1, \om_2)$ obtains current densities,
\beq{}
 j^\mu(\om) = \sum_{\nu, \rho = x, y} \sigma^{\mu \nu \rho}(\om; \om_1, \om_2) E_\nu (\om_1) E_\rho(\om_2)
\eec
as a function of the combined frequencies, $\om = \om_1 + \om_2$, of a pair of optical fields $E_\nu (\om_1), E_\rho(\om_2)$. In the presence of $\mathcal{P}$ symmetry, even-order optical responses generically vanish. Thus, in a centrosymmetric system, the third-order response would be the next nonvanishing higher-order response. To obtain a nonvanishing second-order response, we can break $\mathcal{P}$ symmetry by adding parity-odd diagonal terms, which results in a modified Hamiltonian:
\beq{PB_h}
    h_\mathrm{PB}(\kv)
    =
    h_0(\kv)+t_2\sin{k_x}\mathbbm{1}_3
\eep
We note that the eigenstates of the modified and original Hamiltonians are identical, thus the Euler node topology and the phase diagram remain unaltered. 
The dispersion is globally shifted by $t_2\sin{k_x}$, since $\mathbbm{1}_3=\sum_a P_a(\kv)$, with $P_a(\kv)$ representing a projector onto band $a$. We can then promote the parity-breaking perturbation to the BdG Hamiltonian with a modified form:
\beq{PB_HBdG}
    H_\mathrm{PB}(\kv)
    =
    H(\kv)+t_2\sin{k_x}\mathbbm{1}_6
\eep
As such, the eigenstates of the BdG Hamiltonian under parity breaking term similarly remain unaltered, while the band energy dispersions are globally shifted by $t_2\sin{k_x}$, since $\mathbbm{1}_6=\sum_{a,s} P_{a,s}(\kv)$, with $P_{a,s}$ a projector onto a BdG band. We delegate a further symmetry analysis of the parity effects central to the context of the second-order responses to Appendix~\ref{app::E}.

\begin{figure*}[t]
    \centering
    \includegraphics[width=1\linewidth]{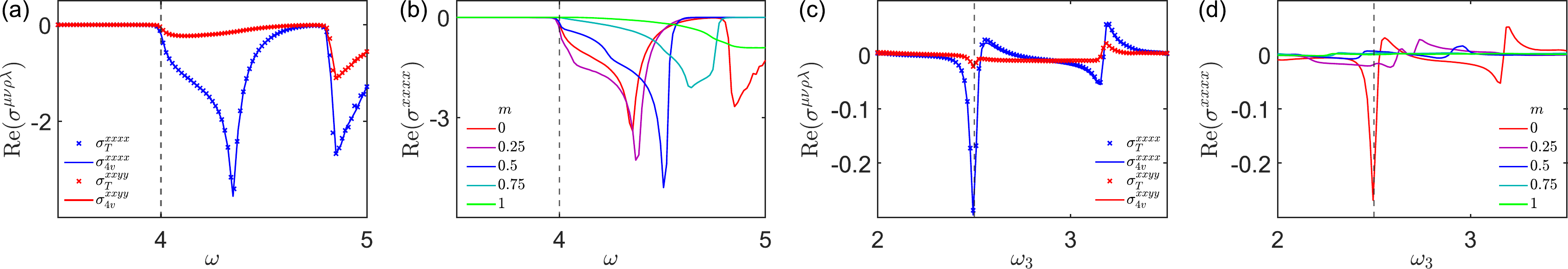}
    \caption{Third-order photoconductivities in the Euler superconductor, with $M=2,\;m=0,\;t=1,\;\D_0=2,\; \eta=10^{-2}$. {\bf (a)} Real part of the total conductivity $\s_T^{\mu\nu\rho\la}(\om;\om,-\om,\om)$, and four-vertex contributions $\s_{4v}$ corresponding to the diagram Fig.~\ref{fig::6}(b) for $\s^{xxxx}$ and $\s^{xxyy}$. We note that the four-vertex contribution is the major contribution. We also observe an increase $\propto\sqrt{\om-2\D_0}$ for $\s^{xxxx}$ for $\om>2\D_0$. {\bf (b)} Similarly to the linear conductivity, the increase persists, provided the Euler node is intact, as we illustrate by varying $m$. {\bf (c)} We further present $\s^{\mu\nu\rho\la}(\om;\om_1,\om_2,\om_3)$, with $\om_2 = 1.5$. Here, $\om_1$ matches the energy difference between the Euler node and the lowest BdG band. The dominant contribution also arises from the four-vertex diagram  and the signature photoconductivity is observed at $\om_2+\om_3=2\D_0$. {\bf (d)} The optical signature gradually dissolves before the Euler node is annihilated, as $m$ is varied in the model. We observe that the photoconductivities with most combinations of spatial indices in $\s^{\mu\nu\rho\la}(\omega;\omega,-\omega,\omega)$ are vanishing, especially, when there are odd numbers of indices in $x$ or $y$. Thus, these negligible components were not shown.}
    \label{fig::5}
\end{figure*}

For responses that are second order in optical fields, it is particularly interesting to address the possible dc current $j^\nu(0)$ responses, that is, in the limit $\om \rightarrow 0$. In particular, these are captured by the photoconductivities $\s^{\mu\nu\nu}(0;\om_1,-\om_1)$. We note that amongst such photoconductivities, the main contribution in the model Euler superconductor is due to the injection currents~\cite{watanabe2022}, with conductivity proportional to the quantum metric:
\beq{con_inj}
    \s_{\text{inj}}^{\mu\nu\rho}(\om) = -\frac{\pi}{4\eta}\sum_{a\neq b}
    (J^\mu_{aa}-J^\mu_{bb})g_{ba}^{\nu\rho}f_{ab}\d(\om-E_{ba})
\eec
and with $\eta$ being the photoexcitation relaxation rate. Here, the imaginary part of the photoconductivity remains vanishing, and the multiband metric can be related to the generalized velocity operators via current operators ${g^{\nu\rho}_{ab}=\Re[J^\nu_{ab}J^\rho_{ba}/E_{ab}^2]}$.

However, within the possible injection current responses, we observe no robust signature of the Euler class, as the intersector multiband metric between Euler bands is vanishing exactly at the node; see Appendix~\ref{app::F}.
We note that the intersector injection current with $\om\sim 2\D_0$ can be separated into two contributions, $\s_{\mathrm{inj};f,-}$ and $\s_{\mathrm{inj};-,-}$, which, respectively, correspond to metrics $g_{f,+;-,-}=g_{f,-;-,+}$ and $g_{-,-;-,+}=g_{-,+;-,-}$. Fundamentally, the latter bears no information about the geometry of the normal state. Instead, it can be related purely to the characteristic band dispersion rather than to the Euler band geometry or topology. 

We demonstrate numerical results for the second-order photoconductivities in Fig.~\ref{fig4}. For $\s^{xxx}$, we compare the case of $\phi = 0$ and $\phi = \pi$ phase differences in Fig.~\ref{fig4}(a). In particular, for $\s^{xxx}$, the major contribution is $\s_{\mathrm{inj};-,-}$, thus, changing $\p$ only affects the contribution that captures the multiband topology and does not cause any significant change in the profile around $\om=2\D_0=4$. We note that the contribution from $\s_{\mathrm{inj};f,-}$ is vanishing when $\p=0$, as the generalized velocity operator $J^\mu_{ab}$ and the multiband metric vanish in that case, leading to the trivial Euler
class.
For $\s^{xyy}$, as illustrated in Fig.~\ref{fig4}(b), we note that $\s_{\mathrm{inj};-,-}$ is now vanishing, since parity is not broken along the $y$ direction. Thus, the major contribution is $\s_{\mathrm{inj};f,-}$. However, since the corresponding metric vanishes exactly at the node, no direct signature of the Euler class can be extracted.
In Figs.~\ref{fig4}(c) and (d), we change $t_2$, which corresponds to the relative strength of the parity-breaking term. We note that both $J^x_{aa}-J^x_{bb}$ and $g^{xx}_{-,-;-,+}$ increase linearly with $t_2$, thus the conductivity increases roughly quadratically for $\s^{xxx}$ and linearly for $\s^{xyy}$.

Furthermore, having studied the second-order responses of the noncentrosymmetric Euler superconductors, we can compute more general third-order responses to the optical fields. The associated current densities read
\begin{small}
\beq{}
j^\mu(\om) = \sum_{\nu, \rho, \lambda = x, y} \sigma^{\mu \nu \rho \lambda}(\om; \om_1, \om_2, \om_3) E_\nu (\om_1) E_\rho(\om_2) E_\lambda (\om_3),
\eeq
\end{small}
with the third-order photoconductivity tensor elements $\sigma^{\mu \nu \rho \lambda}(\om; \om_1, \om_2, \om_3)$ and frequency $\om = \om_1 + \om_2 + \om_3$. 

To fully understand the nature of the third-order photoconductivity analytically, we instead need a diagrammatic approach, which we discuss in the next section; see also Appendix~\ref{app::G}. We here first focus on discussing the numerical result, which we further verify within a diagrammatic approach in the next section. To begin, we investigate the real part of the total conductivity $\s^{xxxx}(\om;\om,-\om,\om)$ in Fig.~\ref{fig::5}(a), which corresponds to four virtual transitions combining the Euler bands between different sectors. We note that $\s^{xxyy}$ is suppressed compared to $\s^{xxxx}$. We also observe that the profile of conductivity $\s^{xxxx}$ against the frequency $\om$ is similar to $\s^{xx}$, with both enjoying a signature at $\om=2\D_0$. To demonstrate the robustness of this correspondence, we vary the value of $m$ in Fig.~\ref{fig::5}(b), observing that the signature remains intact, as long as the Euler node is not gapped out.
However, as we will discuss, we note that the signature is not a discrete jump but a discontinuity scaling as $\sim \sqrt{\om-2\D_0}$. The discontinuity is not quantized, as we further derive in Appendix~\ref{TOC} using the diagrammatic approach.

We also consider the case where $\om_2+\om_3=2\D_0$, where we arbitrarily set $\om_2 = 1.5$. To ensure that the calculated conductivity is mainly contributed by the Euler node, we set $\om_1$, equal to the energy difference between the Euler node, and the lowest band, at the same momentum point. From Fig.~\ref{fig::5}(c), we note that $\s^{xxyy}$ is similarly suppressed when compared with $\s^{xxxx}$, and we note a peak when exactly $\om_2+\om_3=2\D_0$, which we propose as a two-photon transition across the superconducting gap. We do note, however, that as we change $m$ in Fig.~\ref{fig::5}(d), the signature is significantly weakened even when the Euler node is not gapped out. To justify this effect, we remark that the additional bands, which host no Euler class, also contribute to the third-order response.

\section{Diagrammatic analysis}\label{sec::VI}

Previously, we used a response theory approach to calculate the optical conductivity for our specific model. However, such an approach proves difficult to provide clear general insights, linking the Euler class to the response observed numerically.
In the following, we therefore analyze general diagrammatic features within linear and nonlinear response theories~\cite{parker2019}, which pinpoint the nontriviality of the Euler invariant in the superconducting state. We stress that these field-theoretic features are beyond the specific model realizations of the Euler superconductor, see Appendixes~\ref{app::G}-\ref{vc} for full technical details. Given the numerically retrieved anomalous jump originating from the
Euler class in optical conductivity in Sec.~\ref{sec::IV}, we begin by analyzing the associated diagrammatics of the first-order optical conductivity.

\begin{figure}
\centering
\includegraphics[width=0.98\linewidth]{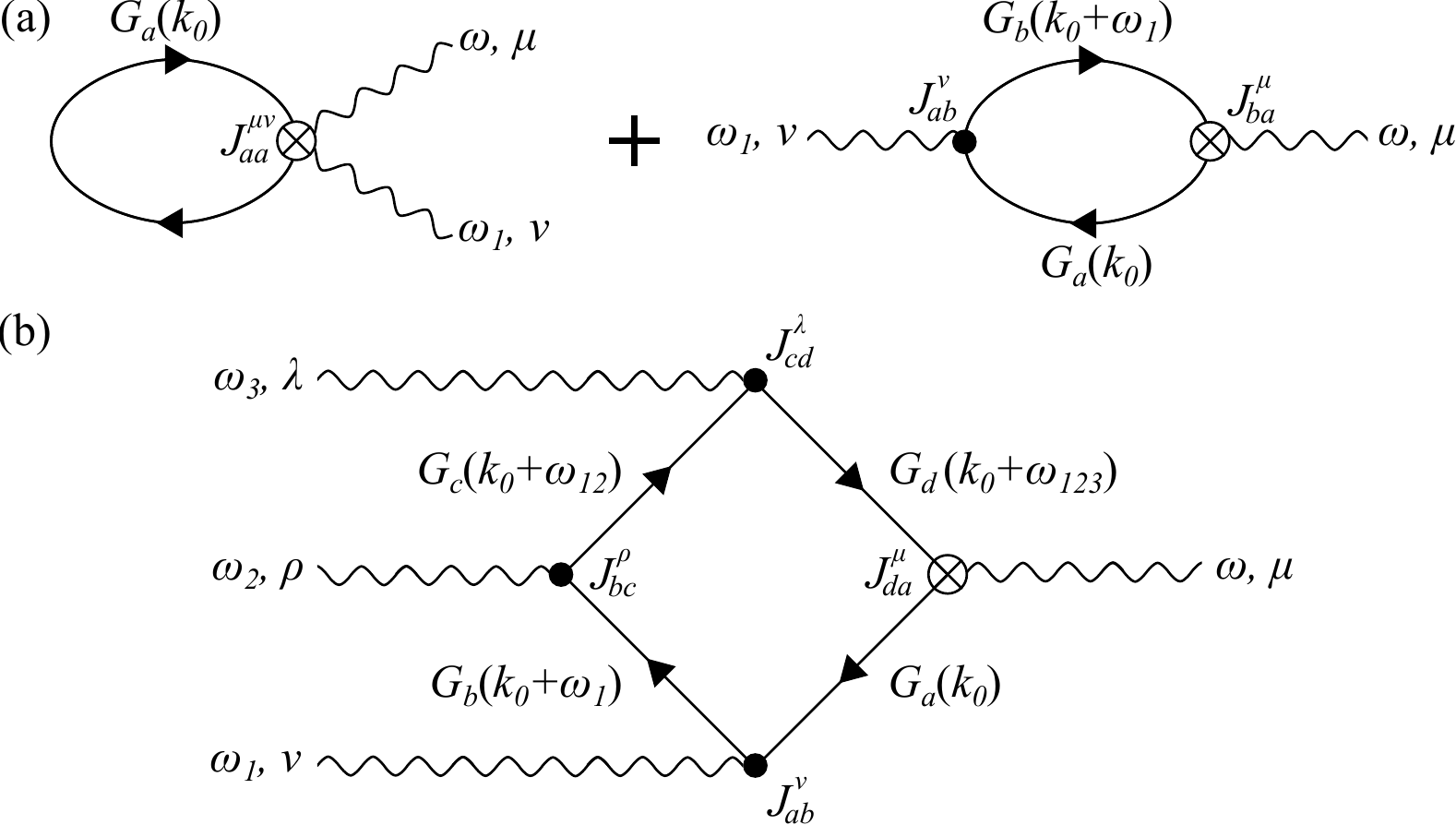}
\caption{One-loop diagrams corresponding to the topologically induced  anomalous optical responses of Euler superconductors. \textbf{(a)} First-order loop diagrams. \textbf{(b)} Third-order loop diagram capturing the enhanced third-order response of an Euler superconductor. Solid dot indicates a vertex with an incoming photon; circle with cross indicates a vertex with an outgoing photon.}
\label{fig::6}
\end{figure}

The total first-order optical conductivity  $\s^{\mu\nu}(\om)$ of a superconductor can be decomposed in terms of two diagrammatic contributions; see Fig.~\ref{fig::6}(a)~\cite{watanabe2024}. Accordingly, expressed in terms of Green's functions $G(k_0,\kv)$ corresponding to the propagators of quasiparticles with four-momenta $k^\mu = (k_0, \kv)$, the conductivity $\s^{\mu\nu}(\om)$ is given by
\begin{align}
    \s^{\mu\nu}(\om)&= \frac{i}{2\om} \int [d\kv]\int dk_0\; \mathrm{Tr}[J^{\mu\nu}(\kv)G(k_0,\kv)
    \nonumber\\
    &+J^\mu(\kv) G(k_0+\om,\kv)J^\nu(\kv) G(k_0,\kv)]\label{lpc},
\end{align}
where $J^{\mu_1\ldots\mu_n}(\kv)$ is the $n$th order generalized velocity operator in the superconducting state \cite{watanabe2022}. The generalized velocity operator of the superconducting state can be related to that of the normal state $j^{\mu_1\ldots\mu_n}(\kv)$, as long as the Hamiltonian is centrosymmetric, i.e., invariant under $\mathcal{P}$ symmetry,
\begin{align}
    j^{\mu_1\ldots\mu_n} (\kv)
    &=\frac{\partial h(\kv)}{\partial k_{\mu_1}\ldots\partial k_{\mu_n}}\; , \\
    J^{\mu_1\ldots \mu_n}(\kv)&=\left.(-1)^n\frac{\partial H(\kv,\A)}{\partial A_{\mu_1}\ldots\partial A_{\mu_n}} \right\vert_{\A\ra 0}
    \nonumber\\
    &= \tau_z^{n+1}\otimes j^{\mu_1\ldots\mu_n}(\kv)\;,
\end{align}
where $\tau_z=\mathrm{diag}(1,-1)$ is the Pauli matrix in Nambu basis. We note that $G^{-1}(k_0,\kv)=k_0-H(\kv)$, which also defines the Green's function of the BdG Hamiltonian, can  be decomposed as a tensor product:
\begin{align}
    G(k_0,\kv)
    &=
    \sum_a
    G_a(k_0,\kv)
    \otimes P_a(\kv)\;,
    \\
    G_a(k_0,\kv)
    &=
    \frac{1}{\om^2_0-\tilde{\ep}_a^2}
    \begin{pmatrix}
        k_0+\ep_a     &   \D_a    \\
        \D_a^*    &   k_0-\ep_a   \\
    \end{pmatrix}\;.
\end{align}
We can now compute the conductivity using Eq.~\eqref{lpc}, in terms of the velocity operator of the normal state
\begin{align}
    \s^{\mu\nu}(\om) &= \frac{i}{2\om}\sum_{a,b}\int [d\kv]\;\big[j_{aa}^{\mu\nu}\d_{ab} I_1 + j^\mu_{ab}j^\nu_{ba}I_2\big]\;,
    \\
    I_1&=\int dk_0 \Tr[\tau_z G_a(k_0)]\;,\\
    I_2&=\int dk_0 \Tr[G_b(k_0+\om)G_a(k_0)]\;,
\end{align}
where $j_{aa}^{\mu\nu}=\partial^\mu\partial^\nu\ep_a$ and $j^\mu_{ab}=\ep_{ab}\xi^\mu_{ab}$, with $\ep_{ab}=\ep_a-\ep_b$ and ${\xi}^\mu_{ab} = i\langle a \vert \partial^\mu b \rangle$. The calculation of the integrals $I_1,\;I_2$ is included within Appendix~\ref{intGF}. We can identify the first diagram as the Drude contribution, which contains no information about the geometry. For the rest of the section, we will calculate only the second diagram, which encodes the information about the quantum geometry of our system. In particular, we are interested in the intersector response, since the intrasector response cannot be efficiently distinguished from the Drude contribution numerically.

Before proceeding, we should recall some features of Euler nodes; see also Appendix~\ref{app::A}. First, the multiband connection is purely imaginary due to the reality of the Hamiltonian, resulting in a vanishing Berry curvature $\Om^{\mu\nu}_{ab}=-2\mathrm{Im}[\xi^\mu_{ab}\xi^\nu_{ba}]$. Furthermore, the connection between the pair of Euler bands is inversely proportional to $r$ around the Euler node, where $r$ is the momentum-space distance from the node. These features have a direct physical consequence, particularly with regard to the real part of the conductivity, as we will discuss later. To proceed with an analytical calculation, we further make the following assumptions: (i)~$|\D_a|=\D_0$ for all bands, (ii)~we work in zero temperature limit, (iii)~$h(\kv)$ is real and invariant under the action of $\mathcal{P}$, $\mathcal{T}$, and $\mathcal{PT}$ symmetries, and (iv)~one of the bands is flat (i.e., $\ep_a=0$) and touches a dispersive band, as known to occur commonly in Lieb and kagome lattices in the material context. At zero temperature, one retrieves only an intersector response, while due to $\mathcal{P}$ symmetry, $\s^{xy}$ vanishes; see Appendix~\ref{app::G}. 

The total response between band indices $a,\;b$, namely, $a,+\leftrightarrow b,-$ and $a,-\leftrightarrow b,+$, is
\begin{small}
\beq{PC_0}
    \s^{\mu\mu}(\om) = \frac{i}{\om}\int [d\kv]\;\ep_{j}^2|\xi_{ab}^\mu|^2
    \frac{(\D_0+\tilde{\ep}_b)
    (\D_0\cos\p-\tilde{\ep}_b)}
    {\tilde{\ep}_b[(\D_0+\tilde{\ep}_b)^2-\om^2]}
\eec
\end{small}
where $\p$ is the phase difference between $\D_a$ and $\D_b$. Despite the fact that the phase difference $\p$ has no effect on the dispersion spectrum of the BdG Hamiltonian, we note that $\p$ has a geometrical consequence, manifested in the multiband connection, by introducing a phase difference  between Bloch band vectors. In particular, when $\om=2\D_0$, which corresponds to the transition between the nodes of electron and hole sectors, $\s^{\mu\mu}\propto 1-\cos\p$,
namely, the response is suppressed when $\p=0$, and maximized when $\p=\pi$. The analytical finding quantitatively agrees with the numerical result in Fig.~\ref{fig3}.

Due to the aforementioned $\xi^\mu_{ab}\propto r^{-1}$ dependency around the Euler node, for each Euler node of integer Euler class $\chi$, constituted by a flat band and another band with dispersion $\ep_j \propto r^{2|\chi|}$, such scalings introduce a jump in $\mathrm{Re}[\s^{\mu\mu}]$, on crossing the gap energy $\om=2\D_0$:
\beq{rPC}
    \mathrm{Re}[\s^{\mu\mu}(\om\ra 2\D_0^+)] = \frac{e^2}{32\hbar}|\chi|(1-\cos\p)
\eec
where we have used $\om\ra 2\D_0^+$ to indicate $\om$ approaching the gap energy $2\D_0$ from above. This precisely corresponds to the optical conductivity jump of $e^2/(16\hbar)$ in the Lieb lattice (see Fig.~\ref{fig3}), as the frequency matches a singular node with Euler class $\chi = 1$, which resides between a flat and a quadratic band when $m=0$.
In particular, when the Euler node is split for $m\neq 0$, each resulting node can be viewed as carrying a patch Euler class of $\chi = 1/2$, and is approximately linear in dispersion. This splitting results in a discrete jump in optical conductivity that is no longer quantized (see Appendix~\ref{App::Robust}), but is geometrically bounded by the total patch Euler class of $\chi = 1$ at $e^2/(8\hbar)$, as discussed further in Appendix~\ref{app::D}. We note that the discrete jump vanishes when the system becomes fully gapped for $m>m_c$, where the total patch Euler class is zero, thus trivializing the geometric bound. To demonstrate that these features are universal, i.e., not specific to particular lattice model realizations, we also retrieve an analogous discrete conductivity jump due to an Euler node in a kagome lattice Hamiltonian, as detailed and illustrated in Appendix~\ref{app::L}.

On the contrary, for second-order conductivity, the results can be well understood within the nonlinear response theory, where the injection current is the major contribution, and no topological terms are acquired. Thus, we leave the discussion of the diagrammatics for the second-order response out of this work.

At third order in optical fields, we find that the dominant contribution arises from a loop diagram introducing four quasiparticle propagators that target the pairs of Euler bands twice; see Fig.~\ref{fig::6}(b). This observation has been verified numerically in Figs.~\ref{fig::5}(a) and (c). We note that the analytical calculation is mostly analogous to the linear case and is therefore detailed in Appendix~\ref{TOC}. From the diagrammatic approach, one can deduce that the individual contributions per momentum space point $\kv$ scale with the square of the Euler curvature. We also note that differently from linear conductivity, where we retrieved a jump originating from the Euler class, the increase of third-order photoconductivity observed in Fig.~\ref{fig::5}(a) is continuous, and $\sigma^{xxxx} \propto\sqrt{\om-2\D_0}$; see also Appendix~\ref{TOC}. Finally, contrary to the normal state~\cite{jankowski2023optical}, we note that all diagrams corresponding to the jerk conductivities~\cite{Fregoso2018} contributed by the transitions only between Euler bands in the particle and hole sectors are suppressed due to a set of third-order selection rules. The third-order selection rules are derived and further detailed in Appendix~\ref{ap_tor}.

\vspace{-10pt}
\section{Discussion}\label{sec::VII}
\vspace{-2pt}

We further comment on the nature and applicability of our findings concerning the optical responses of Euler superconductors. First, it should be noted that anomalous optical conductivity was previously reported in multiorbital superconductors in Ref.~\cite{Chen2021}. Here, we show that a certain class of multiorbital, i.e., multiband, superconductors with nontrivial Euler class moreover inherits an anomalous quantum-geometric first-order photoconductivity response bounded by the total Euler class. We unravelled the topological and geometric character of this anomaly with the diagrammatic perspective in the previous section. In that regard, we also note that within the applied assumptions of intraband interactions, i.e., pairings, the vertex correction remains purely of intraband character. Therefore, the intersector contributions to the linear response are not modified when we introduce vertex corrections, as we also demonstrate in Appendix~\ref{vc}. The vertex corrections contributed by the anomalous propagators admitting hybridized interband pairing channels provide for an interesting future direction to investigate. 

We now discuss the physical picture underlying the anomalous optical conductivity features in the studied Euler superconductors. The jump in the real part and the peaked imaginary part of the first-order conductivity can be linked to the vertex coupling term in diagrammatics shown in Fig.~\ref{fig::6}. The vertex coupling is proportional to the non-Abelian connection $\xi^\mu_{ab}$. The key feature of the Euler node is that $\xi^\mu_{ab}$ is purely imaginary, and from the mathematical point of band geometry, it has a $1/r$ divergence. The divergence leads to a diverging imaginary part of conductivity, which reduces to a finite peak with broadening from $\eta$. For the real part, which vanishes in the absence of divergence, this can be regarded as an anomalous response, as captured by the paramagnetic current bubble term represented by the second diagram in Fig.~\ref{fig::6}(a), which diverges for the quasiparticles at the nodal singularity. Intuitively, the physical meaning of this result is the following. The divergent imaginary part reflects the singular absorption of light by the quasiparticles in the superconductor, which arises because the coupling present in the vertices of the diagrams, and which physically represents interband dipole currents $J^\mu_{ab}$, is enhanced due to topological degeneracy associated with the Euler node. The Euler topology is realized in the orbital basis, as a winding in the non-Abelian Berry connection, and this winding translates into the mixing of orbitals in the eigenstates, which yields enhanced interband dipole moments and associated interband dipole currents $\xi^\mu_{ab}$ on introducing photoexcitations~\cite{jankowski2023optical}. In particular, these interband dipole currents are precisely the vertices of the optical conductivity diagrams capturing photoexcitations in Fig.~\ref{fig::6}. As follows from the Kramers-Kronig causal relations, the absorptive response must be accompanied by a reactive response to light, which gives the jump in the conductivity, when the light couples quasiparticle states with singular interband electric dipole moment induced by the Euler degeneracies. We stress that given the selection rules for optical transitions, our nonvanishing quantization findings hold only specifically for the $d$-wave pairing symmetry. We note that from the physical point of view, the on-site pairing with $\phi = 0$ introduces no interband dipole moment and no interband electric dipole currents, consistent with the vanishing of the vertex coupling. On the contrary, the $d$-wave pairing symmetry of $\phi = \pi$ introduces interband dipole moments in the superconducting state, consistent with the presence of the Euler class, which originally introduced the interband dipole moments in the normal state.

In contrast to the normal Euler phases, such as Euler semimetals, we find distinct effects at the third order in electric fields, namely, the jerk photoconductivity ratios no longer take values inherited from the Euler class~\cite{jankowski2023optical} as soon as the pairing terms are introduced. Instead, we identified a one-loop four-vertex third-order response diagram that provides a topological enhancement rather than a quantization, which can be further traced to the quantum geometry associated with the \text{Euler} \text{invariant}~\cite{Xie2020, Bouhon2023geometric}. 

While both the first- and third-order optical features are generic to the Euler invariant, given the diagrammatic and geometric nature of these anomalies, the second-order responses remain model-dependent.
We showed that the strength of an inversion symmetry-breaking perturbation can significantly alter the response of an Euler superconductor and the injection currents cannot be diagrammatically associated with the nontrivial Euler class. We further note that if the quantizing symmetry, i.e., $\mathcal{C}_2\mathcal{T}$ or $\mathcal{PT}$, is broken, the Euler class in Eq.~\eqref{Eu_Patch} is no longer well-defined, and hence the related quantization of the jump in the first-order optical conductivity is generically no longer preserved. Similarly, one could expect the breakdown of the quantization in the presence of impurity effects as, e.g., modeled with random potential disorder. While we expect that preserving $\mathcal{C}_2\mathcal{T}$ or $\mathcal{PT}$ symmetry on average could preserve the Euler invariant in the renormalized Hamiltonian in principle~\cite{Jankowski2024disorder}, and the anomalous optical signatures could still be retrieved in some form in the presence of disorder, we leave a detailed study of the effects of nonmagnetic and magnetic impurities on Euler superconductors for future work.

Our findings shed light on the experimentally measurable optical responses in Lieb or kagome superconductors, which correspond to the lattices that naturally realize the Euler class invariant. Importantly, a class of cuprate superconductors, in particular, realizing $d$-wave superconducting pairing~\cite{RMPCuprate}, can be modeled with Lieb lattice Hamiltonians~\cite{Yamazaki2020}, which offers for further interesting connections of our optical feature analysis to the exotic correlated states experimentally realizable in these systems. The phase-sensitive character of the optical responses further allows us to optically distinguish Euler superconductors from the other known topological superconductors. $\text{Finally}$, the interplay of the superconducting Euler invariant, quantum geometry, and different possible multiband pairings in the presence of additional bands or disorder, remains an attractive direction for future studies.
\vspace{-10pt}
\section{Conclusion and Outlook}\label{sec::VIII} 
\vspace{-2pt}
We studied anomalous optical signatures of superconductors with a nontrivial multiband Euler class invariant.
We introduced a tunable Lieb lattice model for an Euler superconductor. We found that the superconducting Euler invariant introduces anomalous discrete first-order and enhanced third-order optical responses. We further recast these signatures in terms of diagrammatics at the one-loop order that are applicable to general topological superconductors.
We expect our findings to be relevant to the superconducting states realized on Lieb and kagome lattices, where the nontrivial Euler class can be naturally induced. Our results show that the presence of an exotic multigap topological invariant can be unravelled with anomalous optical responses in multiband superconductors.
\vspace{-10pt}
\begin{acknowledgements}
    \vspace{-2pt}
    The authors thank A.S. Morris for discussions about non-Abelian topologies and superconductivity. C.W.C.~acknowledges funding from the Croucher Cambridge International Scholarship by the Croucher Foundation and the Cambridge Trust.~W.J.J. acknowledges funding from the Rod Smallwood Studentship at Trinity College, Cambridge.~R.-J.S. acknowledges funding from a New Investigator Award, EPSRC Grant No.~EP/W00187X/1, a EPSRC ERC underwrite Grant No.~EP/X025829/1, and Royal Society exchange Grant No.~IES/R1/221060, as well as Trinity College, \text{Cambridge}.
\end{acknowledgements}

\bibliography{references}

\clearpage

\onecolumngrid
\appendix

\section{Euler class and quantum geometry}\label{app::A}

In this appendix, we provide further details on the Euler class as well as on the quantum geometry induced by this multigap topological invariant.

The Euler class is defined with the non-Abelian Euler connection between bands $a$ and $b$~\cite{BJY_nielsen, bouhon2019nonabelian}:
\beq{con}
    \tilde{\xi}^\mu_{ab} = \langle a \vert \partial^\mu b \rangle
\eec
with which we can define the Euler form $\mathrm{Eu}_{ab} = \dd \tilde{\xi}_{ab}$ as an exterior derivative of the connection $1$-form between bands $a$ and $b$:
\beq{Eu_curv}
    \mathrm{Eu}_{ab} = \langle \partial^x a \vert \partial^y b \rangle - \langle \partial^y a \vert \partial^x b \rangle.
\eeq
As mentioned in the main text, the Euler class $\chi_{ab}$ over a two-dimensional BZ patch $\Df$ reads~\cite{BJY_nielsen}
\beq{Eu_Patch2}
    \chi_{ab}(\Df) = 
    \frac{1}{2\pi}\left[
        \int_\Df [d\kv]~\mathrm{Eu}_{ab}(\kv) 
        -
        \oint_{\partial\Df} d\kv \cdot \tilde{\xi}_{ab}
    \right]\;.
\eeq
Centrally to this work, the presence of the Euler invariant induces topologically-enhanced quantum geometry, which drives the optical response of an Euler superconductor. 

The quantum geometry can be captured with the non-Abelian multiband quantum geometric tensor (QGT)~\cite{bouhon2022multigap}, which reads
\beq{}
    (Q_{ab})_{\mu \nu} = \bra{\partial^\mu a} 1 - \hat{P} \ket{\partial^{\nu} b}\;,
\eeq
where $\hat{P} = \sum_{\{c\} }\ket{c}\bra{c}$ is a projector onto a set of bands. In the case of Hamiltonians with the nontrivial Euler invariant $\chi_{ab}$, we choose to project onto the Euler bands:
\beq{}
    \hat{P}_{\chi_{ab}} = \ket{a} \bra{a} + \ket{b} \bra{b}\;.
\eeq
Due to the reality condition on the Hamiltonian $H(\kv) = H^*(\kv)$, the eigenvectors can be chosen real $\ket{a} \in \mathbb{R}^N$, with $N$ given by the number of the orbitals. Hence, the QGT is manifestly real and reduces purely to its real part, the quantum metric $(g_{ab})_{\mu \nu} = \mathrm{Re}~(Q_{ab})_{\mu \nu}$:
\beq{}
    (g_{ab})_{\mu \nu} = \frac{1}{2} \Big( \bra{\partial^\mu a} 1 - \hat{P} \ket{\partial^{\nu} b} + \bra{\partial^\nu b} 1 - \hat{P} \ket{\partial^{\mu} a} \Big)\;.
\eeq
Due to positive-semidefiniteness of QGT~\cite{Xie2020, Bouhon2023geometric, jankowski2023optical}, one arrives at a bound between the quantum metric and the Euler curvature,
\beq{}
    (g_{aa})_{xx} + (g_{bb})_{xx} + (g_{aa})_{yy} + (g_{bb})_{yy} \geq 2|\text{Eu}_{ab}|\;,
\eeq
where we use the projector $\hat{P} = \hat{P}_{\chi_{ab}}$. In the context of this work, the quantum metric decomposes in terms of the interband matrix elements as
\beq{}
    (g_{aa})_{\mu \nu} = \sum_{c \neq b} \xi^\mu_{ac} \xi^\nu_{ca}\;,
\eeq
with the intraband matrix elements $\xi^\mu_{aa} = 0$ vanishing under the reality condition. This allows us to conclude with a geometric bound on the sum of the matrix elements entering the optical conductivities and higher-order responses studied diagrammatically in the following sections:
\beq{}
    \sum_{c \neq b} \xi^x_{ac} \xi^x_{ca} + \sum_{c \neq a} \xi^x_{bc} \xi^x_{cb} + \sum_{c \neq b} \xi^y_{ac} \xi^y_{ca} + \sum_{c \neq a} \xi^y_{bc} \xi^y_{cb} \geq |\text{Eu}_{ab}|\;.
\eeq
In the above, we showed how the Euler curvature provides a lower bound on the matrix elements corresponding to the transitions $from$ and $to$ Euler bands. Furthermore, we can consider the quantum geometry induced by the quadratic band touching hosting patch Euler class $\chi = 1$, and the resultant geometric conditions on the transition $between$ the Euler bands. Within an effective $\kv \cdot \vec{p}$ model for such node, it was shown in Ref.~\cite{jankowski2023optical} that the multiband quantum metric for a rotationally symmetric Euler nodes with patch invariant $\chi$ obtains,
\beq{}
    \xi^x_{ab} \xi^x_{ba} + \xi^y_{ab} \xi^y_{ba} = \frac{\chi^2}{q^4} q^\mu q^\nu (2 \delta_{\mu \nu} - 1)\;,
\eeq
where $q$ is the momentum-space displacement from the position of the Euler node. We note that exactly at the node $q=0$, the quantum metric formally diverges. Furthermore, we show that the presence of such momentum space singularity induces a jump in the first-order optical conductivity (see Appendix~\ref{app::D}), as also discussed in the main text.

\section{Generalized velocity operators}\label{app::C}

Below, for completeness, we show the computation of generalized velocity operator matrix elements in terms of the band dispersion and geometrical quantities~\cite{watanabe2022, watanabe2024}. We begin with the non-Abelian multiband Berry connection
\begin{align}
    \xi^\mu_{ab} &= i\langle a \vert \partial^\mu b\rangle\;,
\end{align}
which determines the matrix elements for the generalized operator ($\mathcal{O}$) derivatives:
\begin{align}
    \mathcal{O}^{\mu}_{ab} &=
     \langle a \vert \partial^\mu \mathcal{O} \vert b \rangle
     \nonumber\\
     &=
     \partial^\mu \langle a \vert  \mathcal{O} \vert b \rangle
     - \langle \partial^\mu a\vert  \mathcal{O} \vert b \rangle
     - \langle  a\vert  \mathcal{O} \vert \partial^\mu b \rangle
     \nonumber\\
     &=
     \partial^\mu \mathcal{O}_{ab} 
     - \sum_c (\langle \partial^\mu a\vert c\rangle\langle c\vert  \mathcal{O} \vert b \rangle
     + \langle  a\vert  \mathcal{O} \vert c\rangle\langle c\vert \partial^\mu b \rangle)
     \nonumber\\
     &=
     \partial^\mu \mathcal{O}_{ab} 
     - \sum_c (i\xi^\mu_{ac}\mathcal{O}_{cb}
     - i \mathcal{O}_{ac} \xi^\mu_{cb})
     \nonumber\\
     &=
     \partial^\mu \mathcal{O}_{ab} 
     - i [\xi^\mu,\mathcal{O}]_{ab}\;.
\end{align}
We first consider the case of a normal state with vanishing pairing terms, as introduced on the Lieb lattice in the main text. For the bulk Hamiltonian, the derivative of which defines the velocity operator, we therefore have
\begin{align}
    h^{\mu}_{ab} &=
    \partial^\mu h_{ab} 
     - i [\xi^\mu,h]_{ab}
     \nonumber\\
     &=
     \partial^\mu \ep_{a}\d_{ab}
     - i\xi^\mu_{ab}\ep_b + i \ep_a \xi^\mu_{ab}
     \nonumber\\
    &=
     \partial^\mu \ep_{a}\d_{ab}
    + i\ep_{ab} \xi^\mu_{ab}\;.
\end{align}
Furthermore, in the superconducting state, the velocity operator can be defined in terms of the velocity operator of the normal state:
\begin{align}
    J^{\mu_1\ldots \mu_n}(\kv)&=\left.(-1)^n\frac{\partial H(\kv,\A)}{\partial A_{\mu_1}\ldots\partial A_{\mu_n}} \right\vert_{\A\ra 0}
    \nonumber\\
    &= \tau_z^{n+1}\otimes h^{\mu_1\ldots\mu_n}(\kv)\;.
\end{align}
Correspondingly, in the band basis of the superconductor model introduced in the main text, we have
\begin{align}
    J^{\mu_1\ldots\mu_n}_{a,s_1;b,s_2}(\kv) &= \langle a,s_1 \vert J^{\mu_1\ldots\mu_n} \vert b,s_2\rangle  
    \nonumber\\
    &=
    \frac{(-1)^{n+1}+\a^*_{a,s_1} \a_{b,s_2}}
    {\sqrt{(1+|\a_{a,s_1}|^2)(1+|\a_{b,s_2}|^2)}}
    h_{ab}^{\mu_1\ldots\mu_n}(\kv)\;,
\end{align}
which provides a convenient representation of the current operators for the diagrammatic calculations performed in the next appendixes.
\section{Diagrammatic approach to the superconducting state}\label{app::G}\label{SC}

We hereby provide a calculation of the optical responses of Euler superconductors within a diagrammatic approach~\cite{watanabe2024}.

The BdG Hamiltonian with $\mathcal{PT}$ symmetry, and specifically, with intraband pairing interactions, can be decomposed as
\beq{BdG_Decompose}
    H(\kv) =
    \sum_a
    \begin{pmatrix}
        \ep_a(\kv)  &   \D_a(\kv)   \\
        \D^*_a(\kv) &   -\ep_a(\kv) \\
    \end{pmatrix}
    \otimes P_a(\kv)
\eec
where $P_a$ is the projector matrix to band $a$ in normal state. Diagrammatically, as shown in Fig.~\ref{fig::6}, the total first-order photoconductivity is given by~\cite{parker2019, watanabe2024}
\beq{PC_1}
    \s^{\mu\nu}(\om) = \frac{i}{\om} \int [d\kv]\int dk_0 \Tr[J^{\mu\nu}G(k_0)+J^\mu G(k_0+\om)J^\nu G(k_0)]
\eep
The Green's function $G^{-1}(k_0,\kv)=k_0-H(\kv)$ can be rewritten as
\begin{align}
    G(k_0,\kv) &= \sum_a 
    \begin{pmatrix}
        k_0 - \ep_a(\kv)  &   -\D_a(\kv)   \\
        -\D^*_a(\kv) &   k_0 + \ep_a(\kv) \\
    \end{pmatrix}
    ^{-1}\otimes P_a
    \nonumber\\
    &=
    \sum_a
    G_a(k_0,\kv)
    \otimes P_a\; ,
    \\
    G_a(k_0,\kv)
    &=
    \frac{1}{k^2_0-\tilde{\ep}_a^2}
    \begin{pmatrix}
        k_0+\ep_a     &   \D_a    \\
        \D_a^*    &   k_0-\ep_a   \\
    \end{pmatrix}\label{GF_SC}.
\end{align}
Furthermore, as the velocity operators can be decomposed as $J^{\mu\nu}=\tau_z\otimes h^{\mu\nu}$ and $J^{\mu}=I_2\otimes h^{\mu}$, we have
\begin{align}
    \s^{\mu\nu}(\om) &= \frac{i}{2\om} \int [d\kv]\int dk_0 \Tr[J^{\mu\nu}G(k_0)+J^\mu G(k_0+\om)J^\nu G(k_0)]
    \nonumber\\
    &=\frac{i}{2\om} \sum_{a,b}\int [d\kv]\int dk_0\; \{h^{\mu\nu}_{aa}\Tr[\tau_zG_a(k_0)]+h^\mu_{ab} h_{ba}^\nu\Tr[G_b(k_0+\om) G_a(k_0)]\}
    \nonumber\\
    &=\frac{i}{2\om} \sum_{a,b}\int [d\kv]\; \big[h^{\mu\nu}_{aa}\tilde{I}_1+h^\mu_{ab} h_{ba}^\nu\tilde{I}_2(\om)\big]\;,
\end{align}
where these integrals are calculated in Appendix~\ref{intGF}. We note that the form is identical to the normal state, with the velocity operator inherited from the normal state, up to the correction that is included within the integral. Moreover, this can be computed explicitly at zero temperature:
\begin{align}
    \frac{i}{\om}\int [d\kv]\; h^{\mu\nu}_{aa}\tilde{I}_1
    &\sim
    -\frac{i}{\om}\int [d\kv]\; \partial^\mu\partial^\nu\ep_a\frac{\ep_a}{\sqrt{\D^2_a+\ep_a^2}}\;,
    \\
    \frac{i}{\om}\int [d\kv]\; h^\mu_{ab} h_{ba}^\nu\tilde{I}_2(\om)
    &\sim
    \frac{i}{\omega}\int[d\mathbf{k}]\;h^\mu_{ab} h_{ba}^\nu
    \frac{(\tilde{\ep}_a+\tilde{\ep}_b)
    (\ep_a\ep_b-\tilde{\ep}_a\tilde{\ep}_b+\D_a\D_b\cos\p)}
    {\tilde{\ep}_a\tilde{\ep}_b[(\tilde{\ep}_a+\tilde{\ep}_b)^2-\om^2]}
    \nonumber\\
    &=
    \frac{i}{\om}\int [d\kv]\; \ep_{ab}^2\xi^\mu_{ab}\xi^\nu_{ba}
    \frac{(\tilde{\ep}_a+\tilde{\ep}_b)
    (\ep_a\ep_b-\tilde{\ep}_a\tilde{\ep}_b+\D_a\D_b\cos\p)}
    {\tilde{\ep}_a\tilde{\ep}_b[(\tilde{\ep}_a+\tilde{\ep}_b)^2-\om^2]}\;.\label{LR_Int}
\end{align} 
\section{Linear response of Euler superconductors}\label{app::D}
Below we detail the calculation for linear response of an Euler superconductor, where we directly apply Eq.~\eqref{LR_Int}. We begin by noting that due to the $C_4$ symmetry, and given the reality condition of the considered Hamiltonians, $\s^{\mu\nu}$ vanishes identically for $\mu\neq\nu$ in the considered models by symmetry in $g^{xy}_{ab}$ (see Appendix~\ref{app::A}), which amounts to zero when integrated over the BZ. As such the non-Drude part of the response, at zero temperature, is given by
\beq{sxx}
    \s_{\text{Euler}} \equiv \frac{\s^{\mu\mu}_{ab}+\s^{\mu\mu}_{ba}}{2}=
    \frac{i}{\om}\int [d\kv]\;\ep_{ab}^2|\xi_{ab}^\mu|^2
    \frac{(\tilde{\ep}_a+\tilde{\ep}_b)
    (\ep_a\ep_b-\tilde{\ep}_a\tilde{\ep}_b+\D_a\D_b\cos\p)}
    {\tilde{\ep}_a\tilde{\ep}_b[(\tilde{\ep}_a+\tilde{\ep}_b)^2-\om^2]}
\eep
We first consider a simplified case where only two bands are optically probed and relevant. We also assume ${\ep_a=0}, {\D_a=\D_b=\D_0}$. Having further assumed no angular dependence for band dispersion, we have the following:
\begin{align}
   \s_{\text{Euler}}&\sim
    \frac{i}{\om A}\int^{r_0}_0\int^{2\pi}_0 rdrd\th\;\frac{(\D_0+\tilde\ep_b)(-\D_0\tilde{\ep}_b+\D_0^2\cos\p)}
    {\D_0\tilde{\ep}_b[(\D_0+\tilde{\ep}_b)^2-\om^2]}\ep_{b}^2|\xi_{ab}^\mu|^2
    \nonumber\\
    &=
    -\frac{i}{\om A}\int^{E_0}_{2\D_0}\int^{2\pi}_0rdEd\th\;
    \lb\frac{d\tilde\ep_b}{dr}\rb^{-1}
    \frac{E[E-\D_0(1+\cos\p)]}
    {\tilde{\ep}_b(E^2-\om^2)}\ep_{b}^2|\xi_{ab}^\mu|^2
    \nonumber\\
    &=
    -\frac{i}{\om A}\int^{E_0}_{2\D_0}\int^{2\pi}_0rdEd\th\; \lb\frac{1}{\ep_b}\frac{d\ep_b}{dr}\rb^{-1}
    \frac{E[E-\D_0(1+\cos\p)]}
    {E^2-\om^2}|\xi_{ab}^\mu|^2
    \nonumber\\
    &\sim  -\frac{i}{\om A}\int^{2\pi}_0d\th\; f^2(\th)\int^{E_0}_{2\D_0} dE\;\frac{r^{2(1+n)}}{m}
    \frac{E[E-\D_0(1+\cos\p)]}{E^2-\om^2}\;,
\end{align}
where $E=\D_0+\tilde{\ep}_b$, $A$ is the area of the BZ, and in the third line, we make an ansatz $|\xi^\mu_{ab}|^2\sim r^{2n}f^2(\th)$ and $\ep_b\sim \a r^m$, locally around the node. The ansatz follows from considering a minimal effective $\vec{k} \cdot \vec{p}$ model for an Euler superconductor,
\beq{}
    H^\chi_{\text{SC}}(\kv) = 
    \begin{pmatrix}
        H^\chi(\kv)& \D_0\mathbbm{1}_2\\
        \D_0\mathbbm{1}_2& -H^\chi(\kv) \\
    \end{pmatrix}
    = \alpha (k^2_x + k^2_y) [\mathbbm{1}_2 \otimes \tau_z] +
    \begin{pmatrix}
        2 \alpha k_x k_y & \alpha(k^2_x - k^2_y) & \D_0 & 0 \\
        \alpha(k^2_x - k^2_y )& -2 \alpha k_x k_y& 0 & \D_0  \\
        \D_0 & 0 & -2 \alpha k_x k_y & \alpha(k^2_y - k^2_x) \\
        0 & \D_0  & \alpha(k^2_y - k^2_x) & 2\alpha k_x k_y
    \end{pmatrix},
\eeq
where $\otimes$ denotes a Kronecker product, and the first term only contributes to the appropriate quadratic or flat band dispersions. The Hamiltonian above corresponds to the case in which the phase difference is $\phi=0$. The nonzero phase difference can be introduced with nontrivial pairing, as detailed in Sec.~\ref{sec::III}. In the above, the nodal Hamiltonian $H^\chi(\kv)$ in the individual particle and hole sectors more compactly reads~\cite{BJY_nielsen, Morris_2024}:
\beq{}
    H^\chi(\kv) = \alpha [(k^2_x + k^2_y) \mathbbm{1}_2 + (k^2_x - k^2_y) \tau_x + 2 k_x k_y \tau_z]
\eep
We note that the form of dispersion controlled by $\alpha$ only contributes to the prefactor of the integral and to the eigenstate normalization factors, which cancel, as $n=-1$ for the quadratic node with Euler class $\chi = 1$ that is centered around $r=0$~\cite{jankowski2023optical}. Specifically in the case of a single Euler node, we retrieve $f(\th)=|\chi|\sin\th$ from the eigenstates~\cite{jankowski2023optical}. Having related the dispersion to nodal Euler class as $m=2|\chi|$~\cite{Morris_2024}, after setting $e = \hbar = 1$, we have:
\begin{align}
    \s_\mathrm{Euler}(\om)&=
    -\frac{i\pi|\chi|}{2\om A}\int^{E_0}_{2\D_0} dE~
    \frac{E[E-\D_0(1+\cos\p)]}{E^2-\om^2},
    \\
    \mathrm{Re}[\s_{\mathrm{Euler}}(2\D_0<\om<E_0)]
    &
    \sim-\frac{i\pi|\chi|}{8\pi^2\om}(\pi i)
    \frac{\om[\om-\D_0(1+\cos\p)]}{2\om}
    \nonumber\\
    &=\frac{|\chi|}{16}
    \lb1-\frac{\D_0}{\om}(1+\cos\p)\rb,
    \label{s_Re}
    \\
    \mathrm{Im}[\s_\mathrm{Euler}(\om)]&=-\frac{\pi|\chi|}{2\om A}\left.
    \mathrm{Re}\lb E-\frac{\om}{2}\ln\frac{\om+E}{\om-E}-\D_0\frac{1+\cos\phi}{2}\ln(E^2-\om^2)\rb
    \right|_{E=2\D_0}^{E_0},
    \nonumber\\
    \mathrm{Im}[\s_\mathrm{Euler}(\om\sim 2\D_0)]
    &\sim
    \left\{
    \begin{aligned}
        &\frac{\pi|\chi|}{8A}(1-\cos\p)\ln|2\D_0-\om|
        &\mathrm{for}\quad \phi\neq0
        \\
        &\frac{\pi|\chi|}{2A}\lb2+2\ln\frac{2\D_0+E_0}{4\D_0}-\frac{E_0}{\D_0}\rb
        &\mathrm{for}\quad \phi=0\;.
    \end{aligned}
    \right.
    \label{s_Im_1}
\end{align}
In the above, we set $A=(2\pi)^2$ when calculating the real part. The real part vanishes outside of $2\D_0<\om<E_0$, and at $\om\sim 2\D_0$ we have a simplified form $\mathrm{Re}[\s_\mathrm{Euler}]=|\chi|{(1-\cos\phi)}/32$ in the units of $e^2/\hbar$. For the imaginary part, at $\om\sim2\D_0$, which corresponds to the transition between Euler nodes in different sectors, there is a divergent peak, except when $\phi=0$.

Below we further check with the exact form of an eigensystem of our concrete model (see Sec.~\ref{sec::II}). When the third band is isolated, only the transition between the flat band to the neighboring dispersive band could give rise to a response. We thus have $\ep_a=0,\;\ep_b=\ep_-,\;\tilde{\ep}_a=\D_0,\;\tilde{\ep}_b=\sqrt{\D_0^2+\ep_-^2}, \;|\xi_{ab}|^2=(\sin^2\th,\cos^2\th)/(r^2+\ep_-^2)$. Substituting back into Eq.~\eqref{sxx} gives:

\begin{align}
    \sigma_\text{Euler}&=\frac{i\pi}{\om A}\int^{r_0}_0 rdr~ \frac{\ep_-^2}{r^2+\ep_-^2}
    \frac{\lb\D_0+\sqrt{\D_0^2+\ep_-^2}\rb
    \lb-\D_0\sqrt{\D_0^2+\ep_-^2}+\D_0^2\cos\p\rb}
    {\D_0\sqrt{\D_0^2+\ep_-^2}\left[\lb\D_0+\sqrt{\D_0^2+\ep_-^2}\rb^2-\om^2\right]}\nonumber\\
    &=
    -\frac{i\pi}{2\om A}\int^{E_0}_{2\D_0} dE~ \frac{E[E-\D_0(1+\cos\p)]}{E^2-\om^2},
\end{align}
where $E_0=\D_0+\sqrt{\D_0^2+\ep^2_-(r_0)}$. Note the function form is identical to that of Euler node with $\chi=1$, thus result can be retrieved by substituting $\chi=1$ into Eqs.~\eqref{s_Re} and \eqref{s_Im_1}. In particular, we note the signature optical conductivity jump of $e^2/(16\hbar)$ at $\om\sim 2\D_0$, which was identified and discussed in the main text for $\p=\pi$.
\\

{
We now also briefly discuss the case when $\D_a\neq\D_b$, where we instead have a discrete jump at $\om=\D_a+\D_b$. We begin by writing the conductivity for this case:
\beq{}
    \s_\mathrm{Euler}=
    -\frac{i\pi|\chi|}{\om A}\int^{E_0}_{\D_a+\D_b} dE~
    \frac{E[E-\D_b(1+\cos\p)]}{E^2-\om^2}
\eec
which has a real part
\begin{align}
    \mathrm{Re}[\s_{\mathrm{Euler}}(\D_a+\D_b<\om<E_0)]
    &\sim-\frac{i\pi|\chi|}{8\pi^2}(\pi i)\frac{\om[\om-\D_b(1+\cos\p)]}{2\om} 
    \nonumber\\
    &=\frac{|\chi|}{16}\lb1-\frac{\D_b}{\om}(1+\cos\p)\rb,
\end{align}
where the jump at $\om=\D_a+\D_b$ is still $e^2/(16\hbar)$ when ${\phi=\pi}$, similarly to the case where $\D_a=\D_b=\D_0$.
\subsection{Beyond flat band limit}
We now discuss the case where both Euler bands are dispersive, respectively, with dispersion $\ep_\pm = \a_\pm r^{2|\chi|}$, to illustrate that the discrete jump in linear conductivity is a general feature for the Euler node, and is anomalous in nature due to divergence in the geometry which is a consequence of the topology.
Applying Eq.~\eqref{sxx}, the linear optical conductivity is given by
\begin{align}
    \s_\mathrm{Euler}
    &\sim
    \frac{i}{\om}
    \int^{r_0}_0
    dr~
    \frac{(\a_+-\a_-)^2|\chi|^2\D_0(\cos\p-1)}
    {4\D_0^2+2(\a_+^2+\a_-^2)r^{2|\chi|}-\om^2}
    r^{2|\chi|-1}
    \nonumber\\
    &=
    \frac{i|\chi|\D(1-\cos\p)}{16\pi\om}
    \frac{(\a_+-\a_-)^2}
    {\a_+^2+\a_-^2}
    \int^{\g_0}_{4\D_0^2} d\gamma~
    \frac{1}{\g-\om^2}\;,
    \\
    \mathrm{Re}[\s_{\mathrm{Euler}}(\om\ra 2\D_0^+)]
    &\sim
    \frac{i|\chi|\D_0(1-\cos\p)}{16\pi\om}
    \frac{(\a_+-\a_-)^2}
    {\a_+^2+\a_-^2}
    (i\pi)
    \nonumber\\
    &=
    \frac{|\chi|}{32}
    \frac{(\a_+-\a_-)^2}{\a_+^2+\a_-^2}
    (1-\cos\phi)\;.\label{sEu_dis}
\end{align}
In the limit where one of the bands is flat, namely, $|\a_+|$ or $|\a_-|\ra 0$ , we again have a jump of $e^2/(16\hbar)$ for $\phi=\pi$ and $\chi=1$, agreeing with the previous calculation.
Another interesting case appears when $\a_+=-\a_-$, resulting in a system similar to a graphene multilayer \cite{2014PhRvB..89c5405J}, where the pair of Euler bands have equal but opposite dispersion. 
For this case, the jump in the real part of the linear conductivity is maximized and given by $|\chi| e^2/(8\hbar)$ for $\phi=\pi$. As such, one can obtain a bound on the discrete jump in conductivity. 
Given there are $N$ Euler nodes each with nodal Euler class $\chi_i$ for $i = 1,\;2\ldots N$ at chemical potential $\mu$, we have
\beq{bound_Eu}
    \mathrm{Re}[\s_{\mathrm{Euler}}(\om\ra 2\D_0^+)]
    \leq
    \frac{e^2}{8\hbar}
    \sum^N_{i=1}|\chi_i|
\eec
where we have showcased that the discrete jump is bounded by the total nodal Euler class $\chi_T=\sum_{i=1}^N|\chi_i|$.

\subsection{Effect of longer range hopping}
\begin{figure*}[t]
    \centering
    \includegraphics[width=0.96\linewidth]{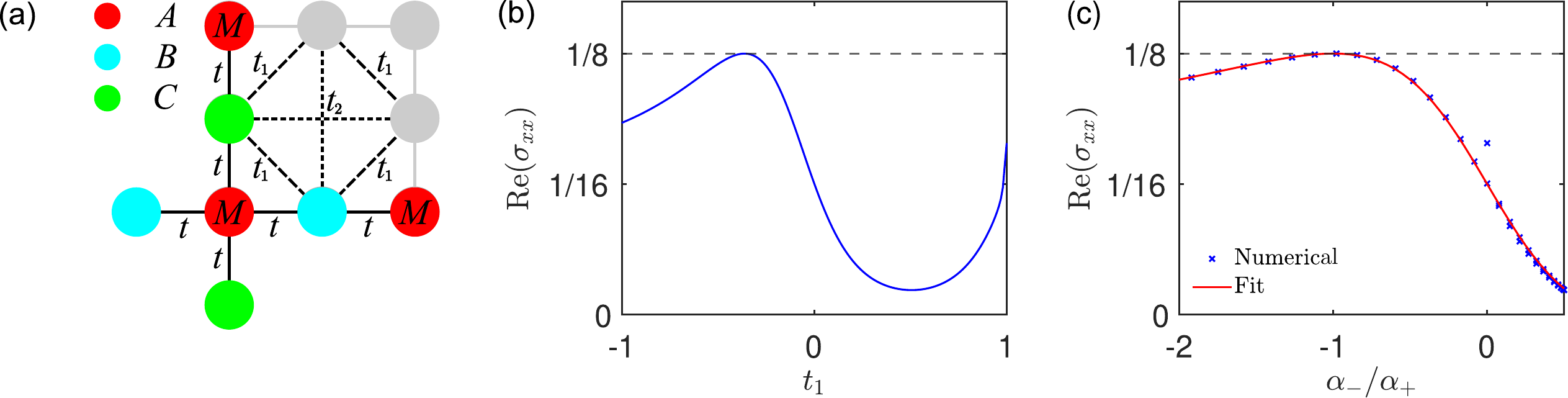}
    \caption{Effect of longer range hopping on linear optical response. {\bf (a)} Lieb lattice with next-nearest and next-next-nearest hopping, respectively $t_1$ and $t_2$. For simplicity, we set the on-site potential for both sites $B$ and $C$ to vanish. {\bf(b)} Plot of discrete jump at $\om=2\D_0$, given chemical potential is set to coincide with the Euler node, as a function of strength of next-nearest-neighbor hopping $t_1$, when $t=1,\;M=2$, and $\D_0=2$. We also set $t_2=-t_1$ such that the Bloch state of the new Hamiltonian including longer range hopping can be related to the original Hamiltonian considered within the main text. We note the maximal value of the jump cannot exceed $e^2/(8\hbar)$, as predicted by the bound Eq.~\eqref{bound_Eu} for total Euler class $\chi_T =1$. {\bf(c)} Using Eqs.~\eqref{nnn_p} and \eqref{nnn_n}, we calculate the ratio of the dispersiveness between the Euler bands as a function of $t_1$, and compute the fit using Eq.~\eqref{sEu_dis}. We note that the jump in conductance is maximized when the two bands have equal but opposite dispersion (namely, $\a_-/\a_+=-1$). We further note that the outlier at $\a_-/\a_+=0$ corresponds to $t_1=1=M/2$, where the system becomes gapless, thus the Euler node is no longer intact.} 
    \label{fig::10}
\end{figure*}
Below we consider the effect of adding longer range hopping, that preserves the symmetry of the lattice. 
In particular, we consider up to next-next-neighbor hopping, as illustrated in \figref{fig::10}(a). 
The Hamiltonian on the Lieb lattice, Eq.~\eqref{HLieb}, including such hopping can be written as
\beq{HLieb_nnn}
    h'(\kv)
    =
    \begin{pmatrix}
        M &   2t\cos\frac{k_x}{2} &   2t\cos\frac{k_y}{2} \\
        2t\cos\frac{k_x}{2} &   m+2t_2\cos k_y   &   4t_1\cos\frac{k_x}{2}\cos\frac{k_y}{2}   \\
        2t\cos\frac{k_y}{2} &   4t_1\cos\frac{k_x}{2}\cos\frac{k_y}{2}   &   -m+2t_2\cos k_x   \\
    \end{pmatrix}
\eec
where we have assumed $m=0$ for simplicity. We note the modified Hamiltonian can be related to the original Hamiltonian, given $t_2=-t_1$, and rewritten as
\beq{}
    h'(\kv)
    =
    h_0(\kv,\;M-2t_1)
    +2t_1\mathbbm{1}_3
    +4t_1\lb
    \cos^2\frac{k_x}{2}
    +
    \cos^2\frac{k_y}{2}
    \rb
    P_f(\kv)
\eep
As such, the longer range hopping has the following effects given the decomposition: (i) on-site energy for site $A$ effectively change from $M$ to $M-2t_1$; (ii) the energy dispersion is universally shifted by $2t_1$ for all bands; and (iii) the flat band becomes dispersive, with the resulting energy dispersion $\epsilon_f = 4t_1[\cos^2(k_x/2)+\cos^2(k_y/2)]$.

We note that the Euler node with $\chi=1$ remains intact when $M\neq 2t_1$. For $M<2t_1$, the pair of Euler bands, around the node, respectively, have dispersion
\begin{align}
    \ep_+ &\sim \frac{t_1}{t^2}r^2
    \label{nnn_p}
    \; ,\\
    \ep_- &\sim -\frac{2}{M-2t_1}r^2
    \label{nnn_n}
    \; ,
\end{align}
which allows us to verify Eq.~\eqref{sEu_dis}. In \figref{fig::10}(b), we plot the discrete jump as a function of $t_1$, and in \figref{fig::10}(c) as a function of $\a_+/\a_-$, which fits with the prediction from Eq.~\eqref{sEu_dis}.
}
\section{Decomposition and analysis of parity effects on the second-order optical responses}\label{app::E}\label{PB}
We briefly comment on the effects of parity breaking, in particular focusing on the second-order optical responses. We begin by recognizing that, given a Hamiltonian $h(\kv)$, we can decompose it in terms of its even [$e(\kv)$] and odd [$o(\kv)$] parts:
\begin{align}
    h(\kv)&=o(\kv)+e(\kv)\;,\\
    o(-\kv)&=-o(\kv)\;,\\
    e(-\kv)&=e(\kv)\;.
\end{align}
We now employ a parity-based decomposition to the generalized velocity operators (see also Appendix~\ref{app::C}), which obtains
\begin{align}
    J^{\mu_{1}\ldots\mu_{n}}(\kv)&=
    \left.(-1)^{n}\prod_{i=1}^{n}
    \frac{\partial}{\partial A^{\mu_{i}}}
    H(\kv,\boldsymbol{A})\right|_{\boldsymbol{A}\rightarrow0}
    \nonumber\\
    &=\tau_{3}^{n}
    \begin{pmatrix}
        j(\kv) & 0\\
        0 & -j(-\kv)
    \end{pmatrix}
    \nonumber\\
    &=\tau_{3}^{n}\left[
    \begin{pmatrix}
    j_{e}^{\mu_{1}\ldots\mu_{n}}(\kv) & 0\\
    0 & -j_{e}^{\mu_{1}\ldots\mu_{n}}(\kv)
    \end{pmatrix}+
    \begin{pmatrix}
    j_{o}^{\mu_{1}\ldots\mu_{n}}(\kv) & 0\\
    0 & j_{o}^{\mu_{1}\ldots\mu_{n}}(\kv)
    \end{pmatrix}\right]\\
    &=\left\{
    \begin{aligned}
    \tau_{3}\otimes j_{e}^{\mu_{1}\ldots\mu_{n}}+\mathbbm{1}_{2}\otimes j_{o}^{\mu_{1}\ldots\mu_{n}} &\quad n\mathrm{\;is\;even}\\
    \tau_{3}\otimes j_{o}^{\mu_{1}\ldots\mu_{n}}+\mathbbm{1}_{2}\otimes j_{e}^{\mu_{1}\ldots\mu_{n}} &\quad n\mathrm{\;is\;odd}
    \end{aligned}\;,
\right.
\end{align}
where we define $j^\mu_{o}\equiv\langle a\vert\partial^{\mu}o\vert b\rangle$ and $j^\mu_{e}\equiv\langle a\vert\partial^{\mu}e\vert b\rangle$. Thus, in the BdG  band basis, we have:
\begin{align}
    J_{-,-;-,+}^{\mu}
    &=\frac{(\a_{-,-}\a_{-,+}-1)j_{o}^{\mu}+(\a_{-,-}\a_{-,+}+1)j_{e}^{\mu}}
    {\sqrt{(1+|\a_{-,-}|^{2})(1+|\a_{-,+}|^{2})}}
    \nonumber\\
    &=-\frac{\D_{0}}{\sqrt{\D_{0}^{2}+\ep_{e}^{2}}}\partial^{\mu}\ep_{o}\;,
    \\
    J_{-,s;-,s}^{\mu}
    &=\frac{(|\a_{-,s}|^{2}-1)j_{o}^{\mu}+(|\a_{-,s}|^{2}+1)j_{e}^{\mu}}
    {1+|\a_{-,s}|^{2}},
    \nonumber\\
    &=j_{e}^{\mu}+\frac{s\ep_{e}}{\sqrt{\D_{0}^{2}+\ep_{e}^{2}}}j_{o}^{\mu}\;,
    \\
    \D_{-,+;-,-}^{\mu}&=\frac{2\ep_{e}}{\sqrt{\D_{0}^{2}+\ep_{e}^{2}}}j_{o}^{\mu}\;,
\end{align}
where $\ep_e$ and $\ep_o$ are the even and odd parts of the dispersion respectively, and $\D^\mu_{-,+;-,-}\equiv J^\mu_{-,+;-,+} - J^\mu_{-,-;-,-}$. Therefore, as the main second-order response is due to the $-,-\ra -,+$ processes, the optical conductivity at the second order in electric fields does not depend on the quantum geometry of the normal state, but only on the dispersion. Hence, unlike the first-order response, the second-order response cannot be directly used to target and infer the topological Euler invariant.

\section{Selection rules for the Euler nodes in superconductors}\label{app::F}\label{sr_sc}
In the following, we discuss an emergent selection rule exactly at the node, which is due to the vertex contribution $J^{\mu_1\ldots\mu_n}_{a,s_1;b,s_2}$. Here, $a,\;b$ are band indices and $s_1,\;s_2=\pm 1$ are sector indices, with $+1\;(-1)$ corresponding to electron (hole) sectors. We note that the eigenstate of the BdG Hamiltonian can be decomposed as a tensor product between the Nambu basis and the band basis:
\begin{align}
    \ket{a;s_1}&= \ket{a,s_1}\otimes \ket{a}\;,\\
    \ket{a,s_1}&= \frac{1}{\sqrt{2}}
    \begin{pmatrix}
        s_1e^{i\p_i}  \\
        1
    \end{pmatrix}\;.
\end{align}
Similarly, we can decompose the coupling at the vertex, see Fig.~\ref{fig::6}, as:
\begin{align}
    J^{\mu_1\ldots\mu_n}_{a,s_1;b,s_2}
    &=
    h^{\mu_1\ldots\mu_n}_{a,b}
    \langle a,s_1\vert \tau_z^{n+1}\vert b,s_2 \rangle
    \nonumber\\
    &=
    \frac{1}{2}
    h^{\mu_1\ldots\mu_n}_{a,b}
    [(-1)^{n+1}+s_1s_2e^{i\D\p}]\;.
\end{align}
Two interesting cases arise when $\D\p=\p_2-\p_1=0,\;\pi$, where certain vertices vanish depending on whether the transition is of inter- or intrasector kind. These selection rules, for the vanishing of the coupling to the current operator at the vertex, can be summarized as:
\\
\begin{equation}
    \begin{tabular}{|c|c|c|c|}
        \hline
        $n$ & $\D\p$ & Sector & Selection rules\\
        \hline
        \multirow{4}{*}{Odd}
        &
        \multirow{2}{*}{0}
        &
        Intra
        &
        \cmark
        \\
        \cline{3-4}
        & &
        Inter
        &
        \xmark
        \\
        \cline{2-4}
        &
        \multirow{2}{*}{$\pi$}
        &
        Intra
        &
        \xmark
        \\
        \cline{3-4}
        & &
        Inter
        &
        \cmark
        \\
        \hline
        \multirow{4}{*}{Even}
        &
        \multirow{2}{*}{0}
        &
        Intra
        &
        \xmark
        \\
        \cline{3-4}
        & &
        Inter
        &
        \cmark
        \\
        \cline{2-4}
        &
        \multirow{2}{*}{$\pi$}
        &
        Intra
        &
        \cmark
        \\
        \cline{3-4}
        & &
        Inter
        &
        \xmark
        \\
        \hline
    \end{tabular}
\end{equation}
\\
Note that the selection rule for band indices $i=j$ is the same as for when $\D\phi=0$.

\section{Third-order conductivity}\label{TOC}

Within a further application of the diagrammatic techniques demonstrated in the approach summarized previously (Appendix~\ref{app::G}), we can deduce the third-order conductivity:
\begin{figure}[h]
    \centering
    \includegraphics[width=0.8\linewidth]{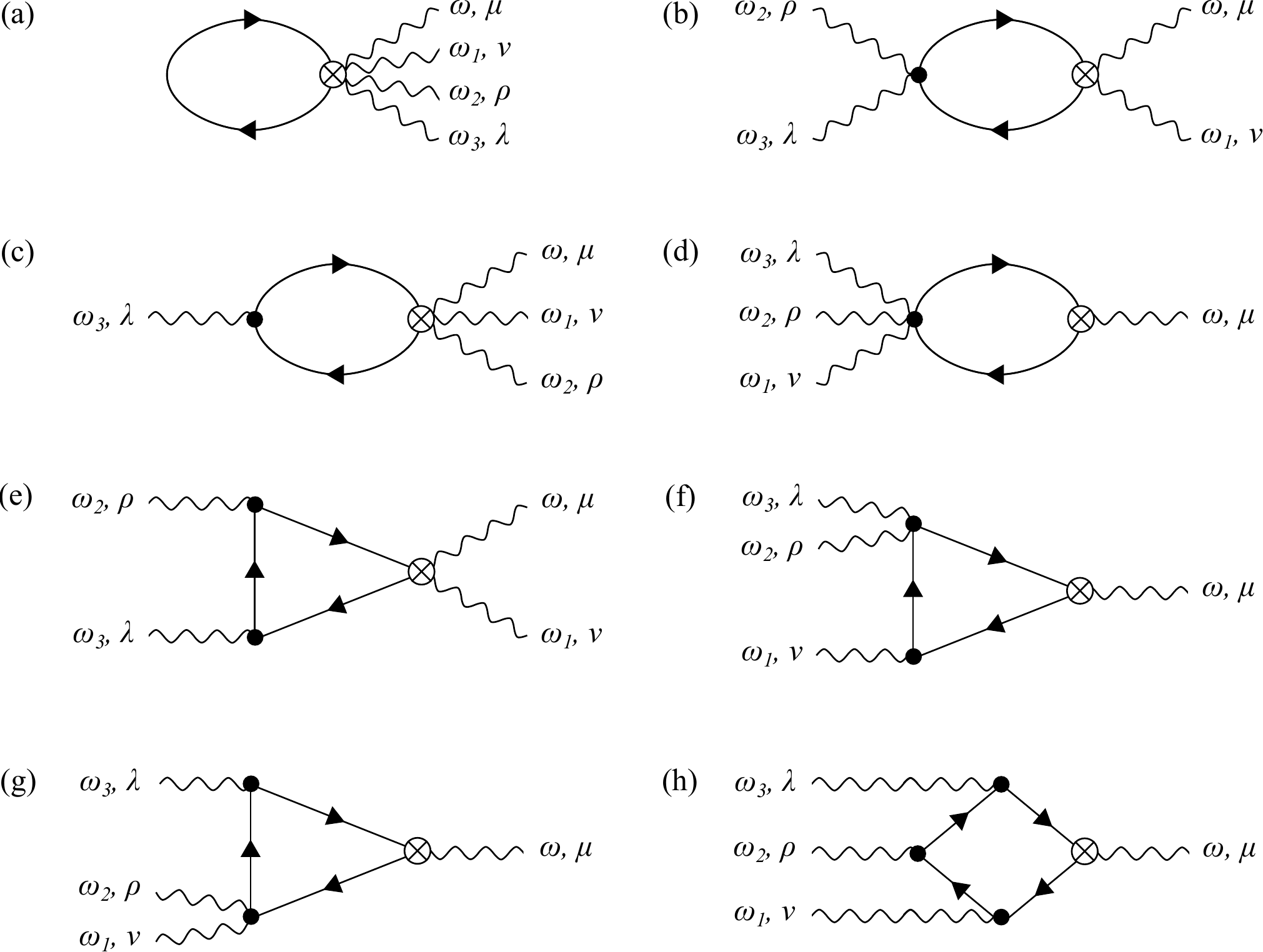}
    \label{fig_diagramatic}
    \caption{List of possible diagrams for third-order optical responses. In the main text, we have only discussed the four-vertex diagram {\bf (h)}, since it is the only diagram that majorly contributes to the optical response of an Euler superconductor.}
    \label{Fig:H}
\end{figure}

\begin{align}
    &\;2\s^{\mu\nu\rho\la}(\om;\om_1,\om_2,\om_3)
    \nonumber\\
    =\;& -\frac{i}{3!\om_1\om_2\om_3}\sum_a\int[d\kv]\int dk_0\;
    G_a(k_0)J_{aa}^{\mu\nu\rho\la}
    \nonumber\\
    & -\frac{i}{2!\om_1\om_2\om_3}\sum_a\int[d\kv]\int dk_0\;
    G_a(k_0)J_{ab}^{\nu}G_b(k_0+\om_1)J_{ba}^{\mu\rho\la}
    \nonumber\\
    & -\frac{i}{2!\om_1\om_2\om_3}\sum_a\int[d\kv]\int dk_0\;
    G_a(k_0)J_{ab}^{\nu\rho}G_b(k_0+\om_{12})J_{ba}^{\mu\la}
    \nonumber\\
    & -\frac{i}{3!\om_1\om_2\om_3}\sum_a\int[d\kv]\int dk_0\;
    G_a(k_0)J_{ab}^{\nu\rho\la}G_b(k_0+\om_{123})J_{ba}^{\mu}
    \nonumber\\
    & -\frac{i}{\om_1\om_2\om_3}\sum_a\int[d\kv]\int dk_0\;
    G_a(k_0)J_{ab}^{\nu}G_b(k_0+\om_{1})J_{bc}^{\rho}G_c(k_0+\om_{12})
    J_{ca}^{\mu\la}
    \nonumber\\
    & -\frac{i}{2!\om_1\om_2\om_3}\sum_a\int[d\kv]\int dk_0\;
    G_a(k_0)J_{ab}^{\nu}G_b(k_0+\om_{1})J_{bc}^{\rho\la}G_c(k_0+\om_{123})
    J_{ca}^{\mu}
    \nonumber\\
    & -\frac{i}{2!\om_1\om_2\om_3}\sum_a\int[d\kv]\int dk_0\;
    G_a(k_0)J_{ab}^{\nu\rho}G_b(k_0+\om_{12})J_{bc}^{\la}G_c(k_0+\om_{123})
    J_{ca}^{\mu}
    \nonumber\\
    & -\frac{i}{\om_1\om_2\om_3}\sum_a\int[d\kv]\int dk_0\;
    G_a(k_0)J_{ab}^{\nu}G_b(k_0+\om_{1})J_{bc}^{\rho}G_c(k_0+\om_{12})
    J_{cd}^{\la}G_d(k_0+\om_{123})J_{da}^{\mu}\;,
\end{align}
which needs to be symmetrized with respect to all permutations $\nu,\;\om_1;\;\rho,\;\om_2;\;\la,\;\om_3$. As discussed in the main text, we are interested in the last term, which corresponds to the diagram Fig.~\ref{Fig:H}(h). In particular, choosing $\phi=\pi$, for $\mu=\nu=\rho=\la=x$ and $\om_1=-\om_2=\om_3=\om$, with $\om>0$, only the following two processes are important due to the selection rules: $f,-\ra -,+\ra f,-\ra -,+\ra f,-$ and $-,-\ra f,+\ra -,-\ra f,+\ra -,-$. The corresponding photoconductivity contributions are given by
\beq{con_4v}
    -\frac{2}{3}\frac{i}{\om^3}\int [d\kv]\int dk_0 \;
    G_{-,-}(k_0)J^x_{-,-;f,+}G_{f,+}(k_0+\om)J^x_{f,+;-,-}
    G_{-,-}(k_0)J^x_{-,-;f,+}G_{f,+}(k_0+\om)J^x_{f,+;-,-}
\eec
where the factor of $2$ is due to two individual processes, and the factor of $1/3$ is due to the symmetrization with respect to all permutations. In particular, on setting the zero temperature limit, this simplifies to the following integral 
\cite{parker2019}:
\begin{align}
    I_4&=\int dk_0 \;G_{-,-}(k_0)G_{f,+}(k_0+\om) G_{-,-}(k_0)G_{f,+}(k_0+\om)
    \nonumber\\
    &=
    \frac{1}{2i\eta(-\tilde\ep_--\D_0+\om+i\eta)(-\tilde\ep_--\D_0+\om+3i\eta)}
    +
    \frac{1}{-2i\eta(-\tilde\ep_--\D_0+\om+i\eta)(-\tilde\ep_--\D_0+\om-i\eta)}
    \nonumber\\
    &=
    \frac{2}{(\tilde\ep_-+\D_0-\om+i\eta)(-\tilde\ep_--\D_0+\om+i\eta)(-\tilde\ep_--\D_0+\om+3i\eta)}\;.
\end{align}
In the context of the photoconductivity contributions specified above, recall that the velocity operator is given by
\beq{}
    J^x_{-,-;f,+}
    =
    \frac{1}{\sqrt{2}}
    \frac{\D_0-\ep_-+\tilde{\ep}_-}
    {\sqrt{\D_0^2+(\ep_--\tilde\ep_-)^2}}
    j_{f-}^{x}(\kv)\;,
\eeq
which results in
\begin{align}
    2\s^{xxxx}&\sim 
    -\frac{2}{3}\frac{i}{\om^3}\int [d\kv]\;  \frac{I_4}{4}\lb\frac{\D_0-\ep_-+\tilde{\ep}_-}
    {\sqrt{\D_0^2+(\ep_--\tilde\ep_-)^2}}
    \rb^4\ep_-^4|\xi_{f-}^{x}|^4
    \nonumber\\
    &\sim 
    -\frac{2}{3}\frac{i}{\om^3A}\int^{r_0}_0\int^{2\pi}_0 rdrd\th\;  2I_4\a^4r^{4m-4}\chi^4\sin^4\th\;.
\end{align}
We further apply the $\vec{k} \cdot \vec{p}$ model-based ansatz: $|\xi^\mu_{ij}|^2\sim r^{-2}\chi^2\sin\th^2$, and $\ep_j\sim \a r^m$ (see Appendix~\ref{app::D}), locally around the node, as also previously detailed in the linear conductivity calculation of Appendix~\ref{SC}. We note that the result depends on the value of $m$ or, equivalently, on the profiles of the band dispersions, and on the quasiparticle relaxation rate $\eta$. For example, for $m=2$, i.e., for a quadratic band, we have $\s^{xxxx}\propto 1/\eta^2$, and 
\begin{align}
    2\s^{xxxx}&\sim 
    -\frac{2}{3}\frac{i}{\om^3A}\int^{r_0}_0\int^{2\pi}_0 rdrd\th\;  \frac{2\a^4r^{4}\chi^4\sin^4\th}{\eta^2(\om-\D_0-\sqrt{\D_0^2+\a^2r^{4}})}
    \nonumber\\
    &\sim 
    -\frac{2}{3}\frac{i}{\om^3A}\int^{\ep_0}_0\int^{2\pi}_0 d\ep d\th\;  \frac{2\a\ep^{2}\chi^4\sin^4\th}{m\eta^2(\om-\D_0-\sqrt{\D_0^2+\ep^2})}\;,
    \nonumber\\
    \mathrm{Re}[\s^{xxxx}]
    &\propto \frac{\chi^4}{\eta^2}\sqrt{\om-2\D_0}
    \quad\text{for $\om>2\D_0$}\;.
\end{align}
We note that distinctly from the linear response, there is a dependence on $\eta\sim 1/\tau$, where $\tau$ is the relaxation time for the photoexcited particle to decay. Consistently with our numerical findings, we thus note that the transition is no longer a discrete jump but instead evolves into a continuous increase.

\section{Useful integrals}\label{intGF}

Below we discuss integrating over the trace of the product of Green's function of the BdG Hamiltonian, with Green's function defined by \eqref{GF_SC}, which is particularly useful in the context of the integrals performed within the diagrammatic approach (Appendix~\ref{app::G}) applied to the considered superconductors~\cite{watanabe2024}. In particular, at finite temperature, the integral can be related to the Matsubara sum of odd frequency $\om_n = (2n+1)\pi/\b$ for $n\in\mathbb{Z}$. For example:
\begin{align}
    \tilde{I}_1 &= \int d\om \Tr[\tau_zG_a(\om)]\nonumber\\
    &= \int d\om \;\frac{2\ep_a}{\om^2-\tilde{\ep}_a^2}\nonumber\\
    &= -\frac{1}{\b}\sum_n \frac{2\ep_a}{\tilde{\ep}_a^2+\om_n^2}\nonumber\\
    &= f(\tilde{\ep}_a)R_1(\tilde{\ep}_a)+f(-\tilde{\ep}_a)R_1(-\tilde{\ep}_a)\nonumber\\
    &= -\frac{\ep_a}{\tilde{\ep}_a}\tanh\frac{\beta\tilde{\ep_a}}{2}\;,
\end{align}
where $R_1$ is the residue of the integrand of $\tilde{I}_1$. Similarly, we have
\begin{align}
    \tilde{I}_2 &= \int d\om \Tr[G_a(\om)G_b(\om+\om_1)]\nonumber\\
    &= \int d\om \;\frac{\D_a^*\D_b+\D_a\D_b^*+2\ep_a\ep_b+2\om(\om+\om_1)}
    {(\om^2-\tilde{\ep}_a^2)[(\om+\om_1)^2-\tilde{\ep}_b^2]}\nonumber\\
    &= 2\int d\om \;\frac{\D_a\D_b\cos\p+\ep_a\ep_b+\om(\om+\om_1)}
    {(\om^2-\tilde{\ep}_a^2)[(\om+\om_1)^2-\tilde{\ep}_b^2]}\;.
\end{align}
Note that $\om_1$ corresponds to the photon (i.e., bosonic) degrees of freedom, thus it is governed by the even Matsubara frequency $2m\pi/\b$ for $m\in\mathbb{Z}$. In the last line, we have also used $\p$ to represent the phase between the order parameter (see Appendix~\ref{app::C}), and $\D_a$ now refers to the modulus of the order parameter. The integral can be analytically continued to sum over the odd Matsubara frequencies $z$:
\begin{align}
    \tilde{I}_2
    &= \frac{2}{\b}\sum_z \frac{\D_a\D_b\cos\p+\ep_a\ep_b-z(z+i\om_1)}
    {(z^2-\tilde{\ep}_a^2)[(z+i\om_1)^2-\tilde{\ep}_b^2]}\nonumber\\
    &= \sum_{z=\pm\tilde{\ep}_a,\pm\tilde{\ep}_b-i\om_1} f(z)R_2(z)\nonumber\\
    &
    = \frac{(\tilde{\ep}_a+\tilde{\ep}_b)
    (\ep_a\ep_b-\tilde{\ep}_a\tilde{\ep}_b+\D_a\D_b\cos\p)}
    {\tilde{\ep}_a\tilde{\ep}_b[(\tilde{\ep}_a+\tilde{\ep}_b)^2-\om_1^2]}
    [f(-\tilde{\ep}_a)+f(-\tilde{\ep}_b)-f(\tilde{\ep}_a)-f(\tilde{\ep}_b)]
    \nonumber\\
    &
    +\frac{(\tilde{\ep}_a-\tilde{\ep}_b)
    (\ep_a\ep_b+\tilde{\ep}_a\tilde{\ep}_b+\D_a\D_b\cos\p)}
    {\tilde{\ep}_a\tilde{\ep}_b[(\tilde{\ep}_a-\tilde{\ep}_b)^2-\om_1^2]}
    [f(\tilde{\ep}_a)-f(-\tilde{\ep}_a)-f(\tilde{\ep}_b)+f(-\tilde{\ep}_b)]\;,
\end{align}
where the first term corresponds to the intersector processes, whereas the second term corresponds to the intrasector processes. In particular, at zero temperature, the second term vanishes. Given that band $a$ is flat, with $\D_a=\D_b=\D_0$, and $\ep_b\ll\D_0$, we have
\begin{align}
    \tilde{I}_2&=\frac{(\D_0+\tilde{\ep}_b)
    (-\tilde{\ep}_b+\D_0\cos\p)}
    {\tilde{\ep}_b[(\D_0+\tilde{\ep}_b)^2-\om_1^2]}
    \nonumber\\
    &=
    \frac{\D_0(1-\cos\p)}{\om_1^2-4\D_0^2}
    +
    \frac{4\D_0^2+\om_1^2-12\D_0^2\cos\p+\om_1^2\cos\p}{4\D_0(\om_1^2-4\D_0^2)^2}\ep_b^2
    +\bigO(\ep_b^3)\;,
\end{align}
while the intrasector contribution vanishes. We note that this form can be further simplified for $\om_1\sim2\D_a$:
\beq{I2simp}
    \tilde{I}_2
    \sim
    \frac{\D_0(1-\cos\p)}{\om_1^2-4\D_0^2}
    +
    \frac{2\D_0(1-\cos\p)}{(\om_1^2-4\D_0^2)^2}\ep_b^2
    +\bigO(\ep_b^3)
\eec
which is proportional to $(1-\cos\p)$ up to the second order in $\ep_b$, consistently with our findings that were numerically retrieved in the main text and, furthermore, analytically derived in the previous sections.

\section{Allowed processes in third-order optical responses of Euler superconductors} \label{ap_tor}

We now detail the selection rules for the allowed third-order processes in the optical responses of the studied Euler superconductors. Based on all possible selection rules, derived analogously to Appendix~\ref{app::F}, we note that only the following transitions have significant contributions, namely, for $\D\p=0$, we have:
\begin{align*}
    1-&\ra2-\ra2-\ra2-\ra1-\\
    1-&\ra2-\ra2-\ra1-\ra1- \mathrm{(three~permutations)}\\
    1-&\ra2-\ra1-\ra1-\ra1- \mathrm{(three~permutations)}\\
    1-&\ra1-\ra1-\ra1-\ra1- \\
    1-&\Ra2+\ra 1+\not\ra 1-\\
    1-&\Ra2+\ra 2+\not\ra 1-\\
    1-&\Ra1+\ra 1+\not\ra 1-\\
    1-&\Ra1+\ra 2+\not\ra 1-\\
    1-&\Ra2+\Ra1-\\
    1-&\Ra1+\Ra1-\\
    1-&\Rra2-\ra1-\\
    1-&\Rra1-\ra1-\\
    1-&\ra1-
\end{align*}
Here, $\ra$ represents a process with a single photon vertex (see Fig.~\ref{Fig:H}), $\Ra$ for a two photon vertex, and $\Rra$ for a three photon vertex [Figs.~\ref{Fig:H}(c) and~\ref{Fig:H}(d)]. In the last line, presented is a process with a single vertex of four photons [Fig.~\ref{Fig:H}(a)]. We denote the suppressed processes with $\not\ra$. Similarly, we can list down the relevant transitions for $\D\p=\pi$:
\begin{align*}
    1-&\ra2+\ra1-\ra2+\ra1- \mathrm{(three~permutations)}\\
    1-&\ra1-\ra1-\ra1-\ra1- \\
    1-&\Ra2-\ra 1+\not\ra 1-\\
    1-&\Ra2-\ra 2-\not\ra 1-\\
    1-&\Ra1+\ra 1+\not\ra 1-\\
    1-&\Ra1+\ra 2-\not\ra 1-\\
    1-&\Ra2-\Ra1-\\
    1-&\Ra1+\Ra1-\\
    1-&\Rra2+\ra1-\\
    1-&\Rra1-\ra1-\\
    1-&\ra1-
\end{align*}
We note that all optical transition processes between the Euler nodes that correspond to the jerk current~\cite{Fregoso2018} are suppressed for both $\D\p=0$ and $\D\p=\pi$.
The jerklike current contributions, i.e., diagrams with two vertices of one photon and one vertex of two photons, with the two-photon transition being between the Euler bands, including a returning transition, is vanishing unless a band other than the bands hosting the Euler class is involved.

\section{Vertex correction}\label{vc}

We hereby comment on the diagrammatic computation of the vertex correction, which is intrinsic to the optical responses of the considered superconductors. The importance of the vertex correction within a diagrammatic approach to the optical responses of superconductors was addressed in Ref.~\cite{watanabe2024}, where the calculation was performed in the sublattice/orbital basis. 
Below, we will follow an analogous formalism but instead use the band basis, which allows for a simpler decomposition. 
More specifically, we have an interaction Hamiltonian of form $H_\mathrm{int}=-\sum_l U_l\sum_{\kv,\kv'}c_l^\dagger(\kv') c_l^\dagger(-\kv') c_l(-\kv) c_l(\kv)$, which includes purely intraband pairing channels. Note that we can write the interaction in the orbital basis, which is the starting point within the tight-binding approach, and then project to the band basis. However, this would cause complications, both in additional form factors and possible interband pairing interactions, which would partly disobey the decomposition that we used.
The order parameter can be related to the Green's function as
\begin{align}
    \D_{i,\A}(x,y) &= 
    -\frac{1}{2}V(x-y)\Tr [\tau_i G_\A (x,y)]\nonumber\\
     &= -\frac{1}{2}\int [dp]\int[dq]\; e^{-i(p+q)\cdot(x-y)} V(p)\Tr [\tau_i G_\A (q)]\;,\nonumber\\
     \D_{i,\A}(k) &= 
     \int [dz]\; e^{ik\cdot z}\D_{i,\A}(z)
     \nonumber\\
     &= 
     -\frac{1}{2}\int [dp]\int[dq]\int [dz]\; e^{i(k-p-q)\cdot z}  V(p)\Tr [\tau_i G_\A (q)]\nonumber\\
     &= 
     -\frac{1}{2}\int [dp]\int[dq]\; (2\pi)^2\d(k-p-q) V(p)\Tr [\tau_i G_\A (q)]\nonumber\\
     &= 
     -\frac{1}{2}\int[dp]\;  V(k-p)\Tr [\tau_i G_\A (p)]\;,
\end{align}
where $i=x$ corresponds to the real part, and $i=y$ corresponds to the imaginary part. The vertex correction in a~single band is accordingly given by
\begin{align}
    \La^\mu_i(x,y,z)
    &= -\frac{\d \D_{i,\A}(x,y)}{\d A_\mu(z)}
    \nonumber\\
    &=
    \frac{1}{2}V(x-y)\Tr 
    \left[\tau_i \frac{\d G_\A (x,y)}{\d A_\mu(z)}\right]
    \nonumber\\
    &=
    -\frac{1}{2}V(x-y)\Tr 
    \left[\tau_i G_\A (x,y)\frac{\d G^{-1}_\A (x,y)}{\d A_\mu(z)}G_\A (x,y)\right],\;
    \nonumber\\
    \La^\mu_i(k,\om) &= -\int [dp]~ \frac{V(k-p)}{2}
    \Tr\left[
    \tau_i G(p_0+\om,\vec{p})
    \G^\mu(p,\om)G(p)
    \right]\;
    \nonumber\\
    &= -\int [dp]~ \frac{V(k-p)}{2}
    \Tr\left[
    \tau_i G(p+\om)\times
    \lb\g^\mu(p)+\sum_j\La^\mu_j(p,\om)\tau_j \rb
    G(p) \right]\;,
    \label{vc0}
\end{align}

which can be converted to the multiband case by a substitution $\tau_i\rightarrow \tau_i \otimes E_l$ and $V(k-p)\rightarrow V_l(k-p)$, where $[E_l]_{ab} = \bra{a}{P_l}\ket{b}$, with $a,\;b$ being the orbital indices and $u_l$ being the Bloch vector with a band index $l$ in the normal phase. We also note that $G^{-1}(k)=k_0-\HBdG(\kv)$, which when assuming a decomposition $\HBdG(\kv)=\sum_l M_l(\kv)\otimes P_l(\kv)$, takes the form
\begin{align}
    M_l(\kv) &= \begin{pmatrix}
        \ep_l(\kv)   &   \D_l(\kv)    \\
        \D_l^\dagger(\kv)    &   -\ep_l(\kv)  \\
    \end{pmatrix}\;,   \\
    G(k)&=\sum_l(k_0 I_2-M_l)^{-1}\otimes P_l
    \nonumber\\
    &=
    -\sum_l
    \frac{1}{\vert\D_l\vert^2+\ep^2_l-k^2_0}
    \begin{pmatrix}
        k_0+\ep_l  &   \D_l    \\
        \D_l^\dagger    &   k_0-\ep_l   \\
    \end{pmatrix}
    \otimes P_l\;.
\end{align}
Importantly, this is only true if the parity remains intact and if we only have intraband interactions.
A quick sanity check for the order parameter with $V(k-p)=g$, and for a single band obtains
\begin{align}
    \D &= g\int[d\vp]\int^\infty_{-\infty} dp_0 \;
    \frac{\D}{|\D|^2+\ep^2-p_0^2}\nonumber\\
    &= \frac{g}{2}\int[d\vp]~ \frac{\D}{\ept}
    [f(-\ept)-f(\ept)]\nonumber\\
    &= \frac{g}{2}\int[d\vp]~ \frac{\D}{\ept}\tanh\frac{\beta\ept}{2}\;.
\end{align}
Furthermore, assuming that $\D_l$ has no $\kv$-dependence, we simply have $V_l(k-p)=g_l$, and thus $\La^\mu_i(k,i\Om)$ drops the $\kv$-dependence. We note that $\La^\mu_{i,l}$ is nothing but an index, and so can be extracted out of the trace. Then Eq.~$\eqref{vc0}$ becomes 
\begin{align}
    \La^\mu_{i,l}(\om) &= -\frac{g_l}{2}\int [dp]\; 
    \Tr\left[
    \tau_i E_l G(p+\om)\times
    \lb\g^\mu(p)+\sum_{j,l'}\La^\mu_{j,l'}(\om)\tau_j E_{l'} \rb
    G(p) \right]\nonumber\\
    \frac{2}{g_l}\La^\mu_{i,l}(\om) &=
    -\int [dp]\; \Tr\left[
    \tau_i E_l G(p+\om)\g^\mu(\vp)
    G(p) \right]
    -\sum_{j,l'}\La^\mu_{j,l'}
    \int [dp]\; 
    \Tr\left[
    \tau_i E_l G(p+\om)
    \tau_j E_{l'}
    G(p)
    \right]
    \nonumber\\
    Q^\mu_{i,l}(\om)
    &=
    \sum_{j,l'}
    \lb
    \frac{2}{g_l}\d_{i,l;j,l'}-Q_{i,l;j,l'}(\om)
    \rb
    \La^\mu_{j,l'}(\om)
    \;,
\end{align}
where we introduce terms that are defined as
\begin{align}
    Q^\mu_{i,l}(\om)
    &=
    -\int [dp]\; \Tr\left[
    \tau_i E_l G(p+\om)\g^\mu(\vp)
    G(p) \right]
    \nonumber\\
    &=
    -\sum_{l'}\int [dp]\; \Tr\left[
    G_{l'}(p)\tau_i G_l(p+\om)\g_{l,l'}^\mu(\vp)
    \right]\;,\\
    Q^\mu_{x,l}(\om)&=
    -\sum_{l'}\int [dp]\; 
    \frac{\D_{l'}(p_0+\om+\ep_{l})
    +\D_{l}^*(p_0+\ep_{l'})
    +\D_l(p_0-\ep_{l'})
    +\D_{l'}^*(p_0+\om-\ep_{l})}
    {(\vert\D_{l'}\vert^2+\ep^2_{l'}-p^2_0)
    [\vert\D_{l}\vert^2+\ep^2_{l}-(p_0+\om)^2]}
    \Tr[P_{l'}(\vp)P_l(\vp)h^\mu_{l,l'}(\vp)]
    \nonumber\\
    &=
    -2\sum_{l'}\int [d\vp]\; 
    \{I_1(l,l')(\D_{x,l}+\D_{x,l'})
    +I_0(l,l')[\om\D_{x,l'}+i(\ep_l\D_{y,l'}-\ep_{l'}\D_{y,l})]\}
    h^\mu_{l,l'}(\vp)\d_{l,l'}
    \nonumber\\
    &=
    -2\int [d\vp]\; 
    [2\D_{x,l}I_1(l,l)
    +\om\D_{x,l}I_0(l,l)]
    h^\mu_{l,l}(\vp)\;,
    \\
    Q^\mu_{y,l}(\om)&=
    -i\sum_{l'}\int [dp]\; 
    \frac{\D_{l'}(p_0+\om+\ep_{l})
    -\D_{l}^*(p_0+\ep_{l'})
    +\D_l(p_0-\ep_{l'})
    -\D_{l'}^*(p_0+\om-\ep_{l})}
    {(\vert\D_{l'}\vert^2+\ep^2_{l'}-p^2_0)
    [\vert\D_{l}\vert^2+\ep^2_{l}-(p_0+\om)^2]}
    \Tr[P_{l'}(\vp)P_l(\vp)h^\mu_{l,l'}(\vp)]
    \nonumber\\
    &=
    2\sum_{l'}\int [d\vp]\; 
    \{I_1(l,l')(\D_{y,l}+\D_{y,l'})
    +I_0(l,l')[\om\D_{y,l'}-i(\ep_l\D_{x,l'}-\ep_{l'}\D_{x,l})]\}
    h^\mu_{l,l'}(\vp)\d_{l,l'}
    \nonumber\\
    &=
    2\int [d\vp]\; 
    [2\D_{y,l}I_1(l,l)
    +\om\D_{y,l}I_0(l,l)]
    h^\mu_{l,l}(\vp)\;,
    \\
    Q_{i,l;j,l'}(\om)
    &=
    -\int [dp]\; 
    \Tr\left[
    \tau_i E_l G(p+\om)
    \tau_j E_{l'}
    G(p)
    \right]
    \nonumber\\
    &=
    -\int [dp]\; 
    \Tr[\tau_i  G_l(p+\om)\tau_j G_{l'}(p)]
    \Tr[P_lP_{l'}]
    \nonumber\\
    &=
    -\int [dp]\; 
    \Tr[\tau_i  G_l(p+\om)\tau_j G_{l}(p)]
    \d_{l,l'}\;,\\
    Q_{x,l;x,l}(\om)
    &=
    -2\int [d\vp]\; 
    (\D_{x,l}^2-\D_{y,l}^2-\ep^2_l)I_0(l,l)+\om I_1(l,l)+I_2(l,l)
    \;,\\
    Q_{y,l;y,l}(\om)
    &=
    -2\int [d\vp]\; 
    (\D_{y,l}^2- \D_{x,l}^2-\ep^2_l)I_0(l,l)+\om I_1(l,l)+I_2(l,l)
    \;,\\
    Q_{x,l;y,l}(\om)&=
    4\int [d\vp]\; 
    (2\D_{x,l}\D_{y,l}+i\ep_l\om)I_0(l,l)
    \;,\\
    Q_{y,l;x,l}(\om)&=
    4\int [d\vp]\; 
    (2\D_{x,l}\D_{y,l}-i\ep_l\om)I_0(l,l)
    \;,
\end{align}
and where we implicitly used that $\g^\mu_{l,l'}=I_2\otimes h^\mu_{l,l'}$.

Interestingly, in the band basis, all interband terms vanish when we only have an intraband interaction, and all terms contain no information of the quantum geometry of the band. Above, with $\ept_{l}>0$ for clarity, we have defined $I_i$ as
\begin{align}
    I_0(l,l') &= \int dp_0~\frac{1}
    {(\ept_{l'}^2-p_0^2)[\ept_{l}^2-(p_0+\om)^2]}
    \nonumber\\
    &=
    \frac{1}{\beta}\sum_z\frac{1}
    {(\ept_{l'}^2-z^2)[\ept_{l}^2-(z+i\om)^2]}
    \nonumber\\
    &=
    \sum_{z=\pm\ept_{l'},\pm\ept_{l}-i\om}f(z)R_0(z)
    \nonumber\\
    &\ra
    R_0(-\ept_{l'}) + R_0(-\ept_l-\om)
    \nonumber\\
    &=
    \frac{\ept_{l}+\ept_{l'}}{2\ept_{l}\ept_{l'}[(\ept_{l'}+\ept_{l})^2-\om^2]}\;,
    \\
    I_1(l,l') &= \int dp_0~ \frac{p_0}
    {(\ept_{l'}^2-p_0^2)[\ept_{l}^2-(p_0+\om)^2]}
    \nonumber\\
    &=
    \frac{1}{\beta}\sum_{z}\frac{iz}
    {(\ept_{l'}^2+z^2)[\ept_{l}^2-(z+i\om)^2]}
    \nonumber\\
    &=
    \sum_{z=\pm\ept_{l'},\pm\ept_{l}-\om}f(z)R_1(z)
    \nonumber\\
    &\ra
    R_1(-\ept_{l'}) + R_1(-\ept_l-\om)
    \nonumber\\
    &=
    -\frac{\om}{2\ept_{l}[(\ept_{l'}+\ept_{l})^2-\om^2]}\;,
    \\
    I_2(l,l') &= \int dp_0~
    \frac{p_0^2}
    {(\ept_{l'}^2-p_0^2)[\ept_{l}^2-(p_0+\om)^2]}
    \nonumber\\
    &=
    \frac{1}{\beta}\sum_{z}\frac{z^2}
    {(\ept_{l'}^2-z^2)[\ept_{l}^2-(z+i\om)^2]}
    \nonumber\\
    &=
    \sum_{z=\pm\ept_{l'},\pm\ept_{l}-\om}f(z)R_2(z)
    \nonumber\\
    &\ra
    R_2(-\ept_{l'}) + R_2(-\ept_l-\om)
    \nonumber\\
    &=
    \frac{\om^2-\ept_{l}(\ept_{l}+\ept_{l'})}{2\ept_{l}[(\ept_{l'}+\ept_{l})^2-\om^2]}\;,
\end{align}
where $R_i$ denotes the residue of the integrand of $I_i$, and $\om_n$ are the odd Matsubara frequencies. We also take the zero temperature limit to obtain a simple analytic form. Since $\om$ is due to the incoming photon, it has to be an element of the even Matsubara frequencies, thus the corresponding poles in the imaginary axis remain unshifted. With only the $l=l'$ integral being relevant, we list zero temperature results below:
\begin{align}
    I_0(l,l)
    &=\frac{1}{\ept_{l}(4\ept_{l}^2-\om^2)}\;,
    \\
    I_1(l,l)
    &=\frac{\om}{2\ept_{l}(4\ept_{l}^2-\om^2)}\;,
    \\
    I_2(l,l)
    &=\frac{\om^2-2\ept_{l}^2}{2\ept_{l}(4\ept_{l}^2-\om^2)}\;. 
\end{align}
As such, the correlation function can be simplified as
\begin{align}
    Q^\mu_{x,l}(\om)
    &=
    -4\int [d\vp]\; \frac{\om\D_{x,l}}{\ept_l(4\ept_l^2-\om^2)}\partial^\mu \ep_l\;,
    \\
    Q^\mu_{y,l}(\om)
    &=
    4\int [d\vp]\; \frac{\om\D_{y,l}}{\ept_l(4\ept_l^2-\om^2)}\partial^\mu \ep_l\;,
    \\
    Q_{x,l;x,l}&=
    -2\int[d\vp]\;\frac{2(\D_{x,l}^2-\D_{y,l}^2-\ep_l^2)+\om^2+\om^2-2\ept_l^2}{2\ept_l(4\ept_l^2-\om^2)}
    \nonumber\\
    &=
    -2\int[d\vp]\frac{2\D_l^2\cos^2\p-2\ept_l^2+\om^2}{\ept_l(4\ept_l^2-\om^2)}\;,
    \\
    Q_{y,l;y,l}&=
    -2\int[d\vp]\;\frac{2(\D_{y,l}^2-\D_{x,l}^2-\ep_l^2)+\om^2+\om^2-2\ept_l^2}{2\ept_l(4\ept_l^2-\om^2)}
    \nonumber\\
    &=
    -2\int[d\vp]\frac{2\D_l^2\sin^2\p-2\ept_l^2+\om^2}{\ept_l(4\ept_l^2-\om^2)}\;,
    \\
    Q_{x,l;y,l}(\om)&=
    4\int [d\vp]\; 
    \frac{2\D_{x,l}\D_{y,l}+i\ep_l\om}{\ept_l(4\ept_l^2-\om^2)}
    \;,\\
    Q_{y,l;x,l}(\om)&=
    4\int [d\vp]\; 
    \frac{2\D_{x,l}\D_{y,l}-i\ep_l\om}{\ept_l(4\ept_l^2-\om^2)}
    \;.\\
\end{align}
We note that with only the intraband interaction, no information of the quantum geometry is included within the vertex correction. Finally, we note that the vertex correction can be obtained by solving the set of linear equations that read
\begin{align}
    Q^\mu_{x,l}(\om)
    &=
    \lb
    \frac{2}{g_l}-Q_{x,l;x,l}(\om)
    \rb
    \La^\mu_{x,l}(\om)
    -
    Q_{x,l;y,l}(\om)\La^\mu_{y,l}(\om)\;,\\
    Q^\mu_{y,l}(\om)
    &=
    \lb
    \frac{2}{g_l}-Q_{y,l;y,l}(\om)
    \rb
    \La^\mu_{y,l}(\om)
    -
    Q_{y,l;x,l}(\om)\La^\mu_{x,l}(\om)\;.
\end{align}
Vertex correction only affects the transitions of form $(l,\pm)\rightarrow(l,\pm)$ and $(l,\pm)\rightarrow(l,\mp)$. The former correspond to the intraband response, while the latter intersector transitions are forbidden, given that $H^\mu_{l,\pm;l,\mp}=0$. 

We conclude by stressing that there is a fundamental difference between working in the band basis and in the orbital basis. In the orbital basis, even limited to the scope of an on-site interaction only, as soon as we convert to the band basis, a general interaction would be both interband and intraband. This dissimilarity could explain why in Ref.~\cite{watanabe2024} a significant difference due to the band mixing can arise, whereas in our case there is no effect beyond the intraband pairing in the linear optical response of the superconducting state. However, this subtle distinction does have an interesting consequence in the third-order response; namely, another type of transition is now possible, which is because we can have a multiple vertex correction.

\section{Robustness of the quantization in the linear optical conductivity}\label{App::Robust}
\begin{figure*}[h]
    \centering
    \includegraphics[width=0.96\linewidth]{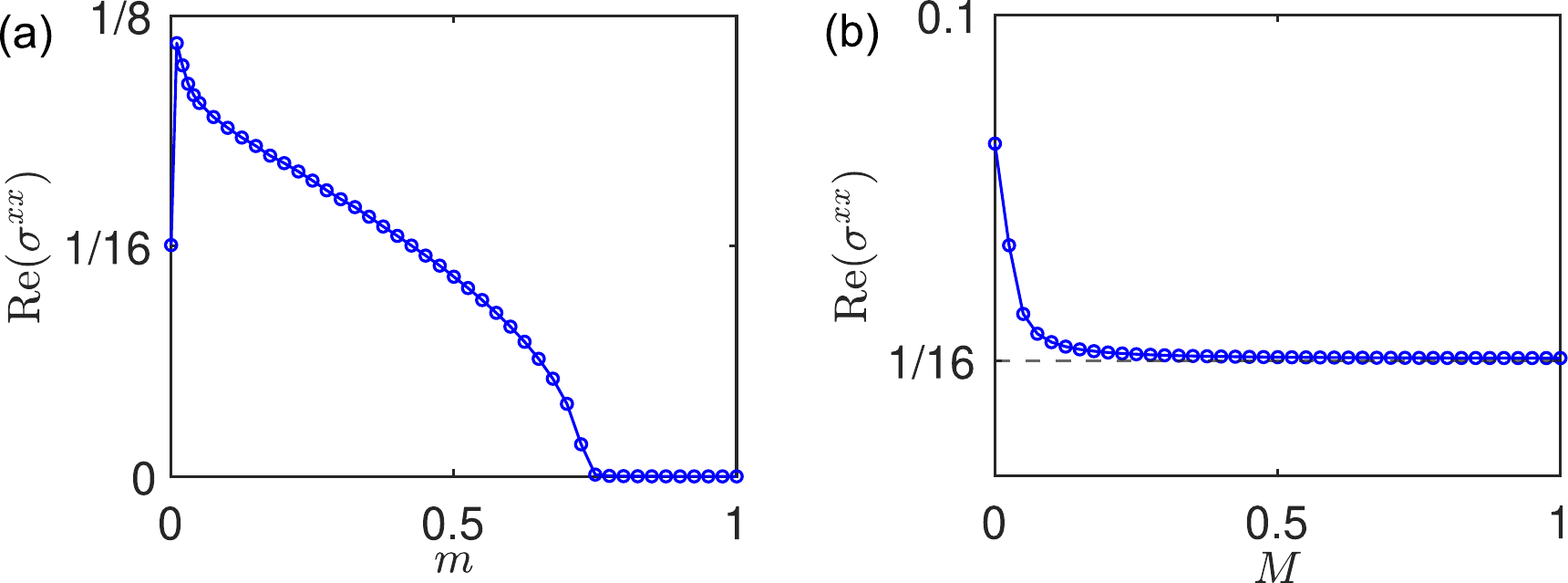}
    \caption{$\mathrm{Re}[\s_{xx}(\om \ra 2\D^+_0)]$ conductance jump as a function of the model parameters $m$ and $M$. {\bf (a)} Conductance jump when $M=2$ and $t=1$, on splitting a quadratic Euler node with patch Euler class $\chi = 1$ into two linear nodes, as $m$ is increased. At $m_c$, cf. Fig.~1(c) of the main text, the nodes annihilate, resulting in $\mathrm{Re}~\s_{xx} = 0$. {\bf (b)} $\mathrm{Re}[\s_{xx}(\om \ra 2\D^+_0)]$ conductivity jump as a function of model parameter $M$. As long as $M$ is sufficiently large, such that the contributions of the other bands to the optical conductivity at $\om \ra 2\D^+_0$ are negligible, the quantization of $\mathrm{Re}[\s_{xx}(\om \ra 2\D^+_0)]$ due to the Euler node is robust.}
    \label{fig::1}
\end{figure*}
In this appendix, we discuss the robustness of the quantization of the linear optical conductivity jump as we change $m$ and $M$.
We note that when $m$ is changed from $m=0$, the Euler node carrying the Euler class $\chi=1$ splits into two linear nodes, which no longer host an integer Euler class individually. Hence, on splitting a single Euler node singularity, as expected, the quantization is broken. However, we do expect the conductance to remain bounded by the total nodal Euler class of $\chi=1$ at $e^2/(8\hbar)$, based on Eq.~\eqref{bound_Eu}. Additionally, we also expect the conductance to vanish when the total nodal Euler class amounts to zero, namely, when the system becomes fully gapped for $m>m_c$. Both of these features are illustrated in \figref{fig::1}(a). 
\\

We do note {that the magnitude of jump originating from the Euler class} persists if the model parameter $M$ is changed instead. This is because both the integer Euler node charge, and the flatness condition of the band remain invariant, given that only the dispersive bands are altered. In \figref{fig::1}(b), we showcase the integer jump approaches the value of $e^2/(16\hbar)$ as $M$ is increased, as the higher band effects on the conductivity become irrelevant.
\color{black}
\newpage

\section{Additional numerical results}\label{app::L}

\begin{figure}[h]
    \centering
    \includegraphics[width=0.8\linewidth]{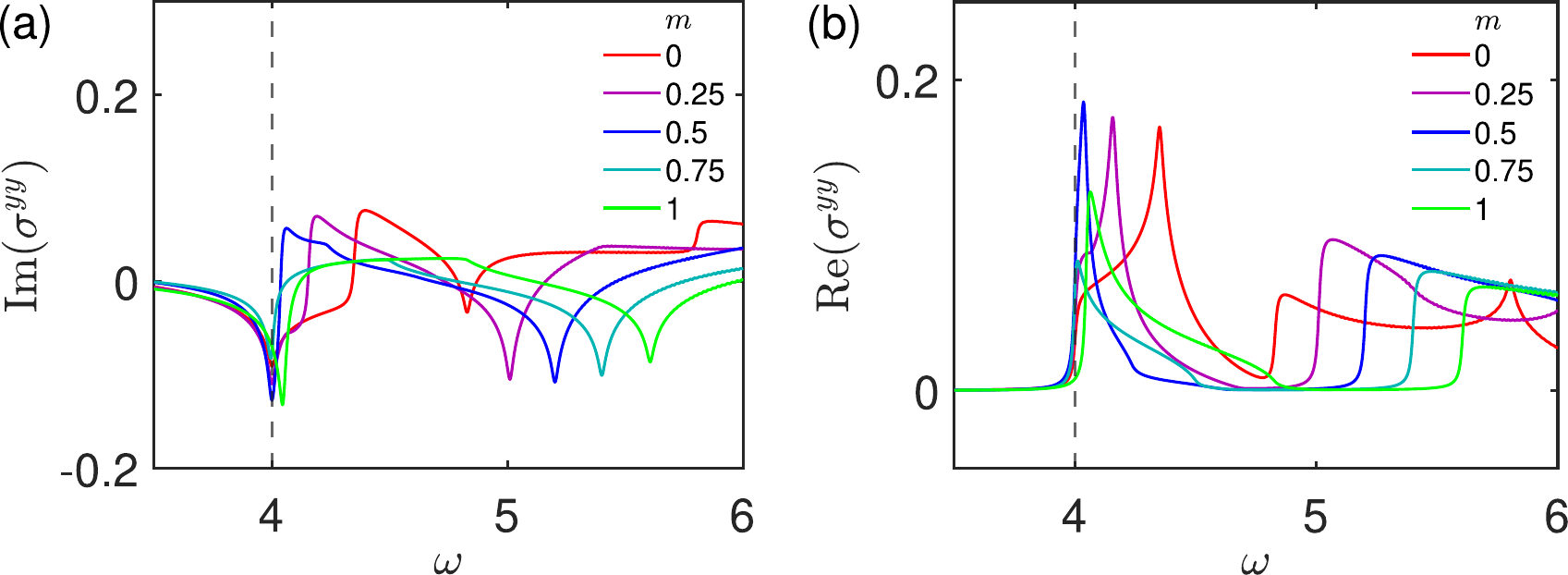}
    
    \caption{Real and imaginary parts of photoconductivity $\s^{yy}$, with the same parametrization as in Fig.~\ref{fig3}. We note that the change in conductivity profile is highly nontrivial when the Euler node is gapped. For example, $\s^{yy}$ can be compared with $\s^{xx}$ in Fig.~\ref{fig3}, accordingly.}
    \label{fig_s_2}
\end{figure}

\begin{figure}[h]
    \centering
    \includegraphics[width=0.8\linewidth]{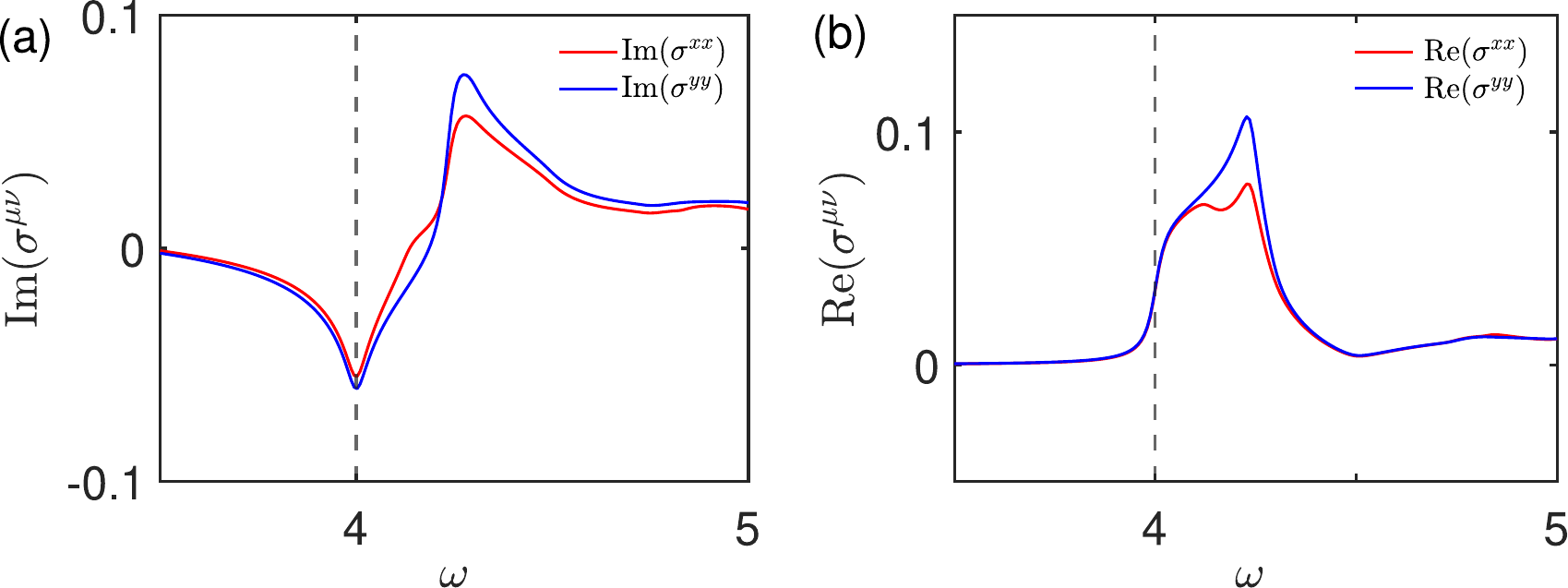}
    
    \caption{Real and imaginary parts of photoconductivities $\s^{xx}$ and $\s^{yy}$, for the superconducting phase realized in a kagome lattice model with $\chi = 1$~\cite{Jiang2021, jankowski2023optical}. We adapt only the nearest-neighbor hopping which is set to $t_1 = 1$, and set all on-site potentials to zero, consistently with the model parametrization of Refs.~\cite{Jiang2021, jankowski2023optical}. For consistency, we set the order parameter in all bands to $\D_0=2$. We note that similarly to the Lieb lattice model, a node of patch Euler class $\chi=1$ is between the flat and quadratic  bands. We set the chemical potential to coincide with the node, similar to Fig.~\ref{fig3}. We observe an analogous conductivity jump in the real part of linear conductivity in the kagome lattice model.}
    \label{fig_s_3}
\end{figure}

\begin{figure}
    \centering
    \includegraphics[width=0.8\linewidth]{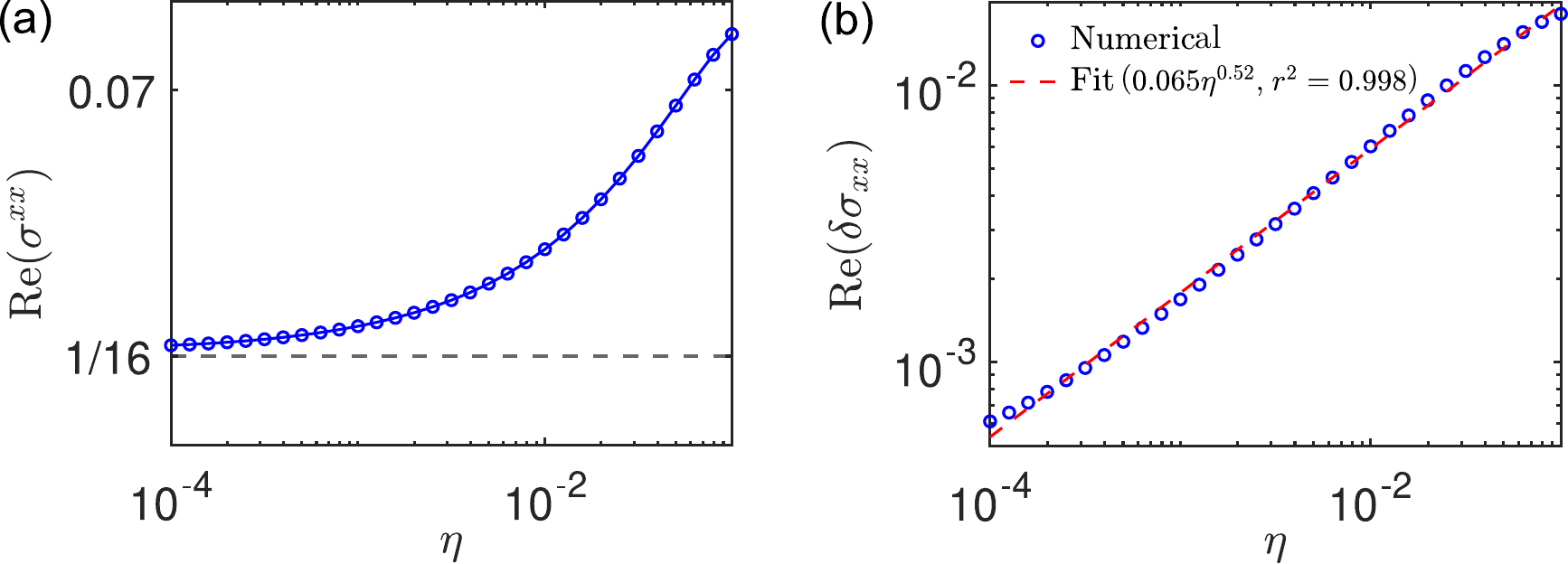}
    \caption{Scaling of the conductivity jump $\mathrm{Re}[\s_{xx}(\om \ra 2\D^+_0)]$ with the numerical broadening parameter $\eta$. {\bf (a)} Conductance jump variation with the numerical parameter $\eta$. {\bf (b)} Numerical error in the conductivity jump $\mathrm{Re}[\d\s_{xx}(\om \ra 2\D^+_0)]$ with respect to the $e^2/(16\hbar)$ value as a function of the grid spacing-dependent numerical parameter $\eta$. The numerical error scales as $\sim\sqrt{\eta}$, as illustrated by the fit.}
    \label{fig:eta}
\end{figure}

In the following, we present further numerical results for the photoconductivities of Euler superconductors realized in the Lieb and kagome lattice models (see Figs.~\ref{fig_s_2} and \ref{fig_s_3}, correspondingly), which were not included in the main text, but for completeness are provided below. In Fig.~\ref{fig:eta}, we demonstrate the scaling of the linear photoconductivity jump in the Lieb lattice model against the numerically set broadening parameter $\eta$.

For reference, the Euler class on a kagome lattice can be implemented by the following Hamiltonian considered and studied in previous works~\cite{Jiang2021, jankowski2023optical}:

\beq{}
    H_\mathrm{kagome}(k_x,k_y)
    =
    \begin{pmatrix}
        E_A&t_1\cos{\frac{k_x}{2}}& t_1\cos{\frac{k_x-\sqrt{3}k_y}{4}}\\
        t_1\cos{\frac{k_x}{2}}& E_B & t_1\cos{\frac{k_x+\sqrt{3}k_y}{4}}\\
        t_1\cos{\frac{k_x-\sqrt{3}k_y}{4}}
        & t_1\cos{\frac{k_x+\sqrt{3}k_y}{4}} & E_C
    \end{pmatrix}
\eec
where $t_1$ is the nearest-neighbor hopping, and $E_A,\ E_B,\ E_C$ are the on-site potentials.

\end{document}